\pdfoutput=1
\documentclass[12pt,a4paper]{article}
\usepackage{ifthen} % for conditional statements
\newboolean{pdflatex}
\setboolean{pdflatex}{true} % False for eps figures 

\newboolean{articletitles}
\setboolean{articletitles}{true} % False removes titles in references

\newboolean{uprightparticles}
\setboolean{uprightparticles}{false} %True for upright particle symbols

\def\paperauthors{LHCb collaboration} % Leave as is for PAPER, CONF and FIGURE
\def\paperasciititle{Search for the doubly charmed baryon Xicc+ in the Xic+Pi-Pi+ final state} % Set ASCII title here !! MAKE sure it's only ASCII characters !! 
\def\papertitle{Search for\\the doubly charmed baryon \Xiccp\\in the \Xicp\pim\pip final state} % Latex formatted title
\def\paperkeywords{{High Energy Physics}, {LHCb}} % Comma separated list
\def\papercopyright{\the\year\ CERN for the benefit of the LHCb collaboration} % new since 9/Apr/2018
\def\paperlicence{CC BY 4.0 licence}
\def\paperlicenceurl{https://creativecommons.org/licenses/by/4.0/}

\usepackage{caption,subcaption}
\usepackage{booktabs}

\usepackage[top=1in, bottom=1.25in, left=1in, right=1in]{geometry}

\usepackage{microtype}
\usepackage{lineno}  % for line numbering during review
\usepackage{xspace} % To avoid problems with missing or double spaces after
\usepackage{caption} %these three command get the figure and table captions automatically small

\usepackage{graphicx}  % to include figures (can also use other packages)
\usepackage{color}
\usepackage{colortbl}
\graphicspath{{./figs/}} % Make Latex search fig subdir for figures
\usepackage{amsmath} % Adds a large collection of math symbols
\usepackage{amssymb}
\usepackage{amsfonts}
\usepackage{upgreek} % Adds in support for greek letters in roman typeset

\newcommand*\patchAmsMathEnvironmentForLineno[1]{%
\expandafter\let\csname old#1\expandafter\endcsname\csname #1\endcsname
\expandafter\let\csname oldend#1\expandafter\endcsname\csname
end#1\endcsname
 \renewenvironment{#1}%
   {\linenomath\csname old#1\endcsname}%
   {\csname oldend#1\endcsname\endlinenomath}%
}
\newcommand*\patchBothAmsMathEnvironmentsForLineno[1]{%
  \patchAmsMathEnvironmentForLineno{#1}%
  \patchAmsMathEnvironmentForLineno{#1*}%
}
\AtBeginDocument{%
\patchBothAmsMathEnvironmentsForLineno{equation}%
\patchBothAmsMathEnvironmentsForLineno{align}%
\patchBothAmsMathEnvironmentsForLineno{flalign}%
\patchBothAmsMathEnvironmentsForLineno{alignat}%
\patchBothAmsMathEnvironmentsForLineno{gather}%
\patchBothAmsMathEnvironmentsForLineno{multline}%
\patchBothAmsMathEnvironmentsForLineno{eqnarray}%
}

\usepackage{hyperxmp}

\usepackage[pdftex,
            pdfauthor={\paperauthors},
            pdftitle={\paperasciititle},
            pdfkeywords={\paperkeywords},
            pdfcopyright={Copyright (C) \papercopyright},
            pdflicenseurl={\paperlicenceurl}]{hyperref}
\usepackage[colorinlistoftodos,textsize=scriptsize]{todonotes}

\usepackage[bottom,flushmargin,hang,multiple]{footmisc}

\usepackage[all]{hypcap} % Internal hyperlinks to floats.

\usepackage{xspace} 
\usepackage{upgreek}

\newcommand{\offsetoverline}[2][0.1em]{\kern #1\overline{\kern -#1 #2}}%

\def\lhcb   {\mbox{LHCb}\xspace}

\def\babar  {\mbox{BaBar}\xspace}
\def\belle  {\mbox{Belle}\xspace}

\def\MagUp {\mbox{\em Mag\kern -0.05em Up}\xspace}

\ifthenelse{\boolean{uprightparticles}}%
{

 \def\Pmu         {\ensuremath{\upmu}\xspace}

 \def\Ppi         {\ensuremath{\uppi}\xspace}                 
                  
 \def\Prho        {\ensuremath{\uprho}\xspace}

 \def\Ppsi        {\ensuremath{\uppsi}\xspace}

 \def\PDelta      {\ensuremath{\Delta}\xspace}                 
 \def\PXi         {\ensuremath{\Xi}\xspace}                 
 \def\PLambda     {\ensuremath{\Lambda}\xspace}                 
 \def\PSigma      {\ensuremath{\Sigma}\xspace}                 
 \def\POmega      {\ensuremath{\Omega}\xspace}                 
 \def\PUpsilon    {\ensuremath{\Upsilon}\xspace}

 \def\PB      {\ensuremath{\mathrm{B}}\xspace}                 
                  
 \def\PD      {\ensuremath{\mathrm{D}}\xspace}

 \def\PJ      {\ensuremath{\mathrm{J}}\xspace}                 
 \def\PK      {\ensuremath{\mathrm{K}}\xspace}

 \def\Pb      {\ensuremath{\mathrm{b}}\xspace}                 
 \def\Pc      {\ensuremath{\mathrm{c}}\xspace}                 
 \def\Pd      {\ensuremath{\mathrm{d}}\xspace}

 \def\Pi      {\ensuremath{\mathrm{i}}\xspace}

 \def\Pp      {\ensuremath{\mathrm{p}}\xspace}

 \def\Ps      {\ensuremath{\mathrm{s}}\xspace}                 
                  
 \def\Pu      {\ensuremath{\mathrm{u}}\xspace}

}
{

 \def\Pmu         {\ensuremath{\mu}\xspace}

 \def\Ppi         {\ensuremath{\pi}\xspace}                 
                  
 \def\Prho        {\ensuremath{\rho}\xspace}

 \def\Ppsi        {\ensuremath{\psi}\xspace}                 
                  
 \mathchardef\PDelta="7101
 \mathchardef\PXi="7104
 \mathchardef\PLambda="7103
 \mathchardef\PSigma="7106
 \mathchardef\POmega="710A
 \mathchardef\PUpsilon="7107
                  
 \def\PB      {\ensuremath{B}\xspace}                 
                  
 \def\PD      {\ensuremath{D}\xspace}

 \def\PJ      {\ensuremath{J}\xspace}                 
 \def\PK      {\ensuremath{K}\xspace}

 \def\Pb      {\ensuremath{b}\xspace}                 
 \def\Pc      {\ensuremath{c}\xspace}                 
 \def\Pd      {\ensuremath{d}\xspace}

 \def\Pi      {\ensuremath{i}\xspace}

 \def\Pp      {\ensuremath{p}\xspace}

 \def\Ps      {\ensuremath{s}\xspace}                 
                  
 \def\Pu      {\ensuremath{u}\xspace}

}

\makeatletter
\ifcase \@ptsize \relax% 10pt
  \newcommand{\miniscule}{\@setfontsize\miniscule{4}{5}}% \tiny: 5/6
\or% 11pt
  \newcommand{\miniscule}{\@setfontsize\miniscule{5}{6}}% \tiny: 6/7
\or% 12pt
  \newcommand{\miniscule}{\@setfontsize\miniscule{5}{6}}% \tiny: 6/7
\fi
\makeatother

\DeclareRobustCommand{\optbar}[1]{\shortstack{{\miniscule (\rule[.5ex]{1.25em}{.18mm})}
  \\ [-.7ex] $#1$}}

\def\mumu       {{\ensuremath{\Pmu^+\Pmu^-}}\xspace}

\def\uquark    {{\ensuremath{\Pu}}\xspace}

\def\dquark    {{\ensuremath{\Pd}}\xspace}
\def\dquarkbar {{\ensuremath{\overline \dquark}}\xspace}

\def\squark    {{\ensuremath{\Ps}}\xspace}

\def\cquark    {{\ensuremath{\Pc}}\xspace}

\def\bquark    {{\ensuremath{\Pb}}\xspace}

\def\pion   {{\ensuremath{\Ppi}}\xspace}

\def\pip    {{\ensuremath{\pion^+}}\xspace}
\def\pim    {{\ensuremath{\pion^-}}\xspace}

\def\rhomeson {{\ensuremath{\Prho}}\xspace}
\def\rhoz     {{\ensuremath{\rhomeson^0}}\xspace}

\def\kaon    {{\ensuremath{\PK}}\xspace}
  \def\Kbar    {{\kern 0.2em\overline{\kern -0.2em \PK}{}}\xspace}

\def\KorKbar {\kern 0.18em\optbar{\kern -0.18em K}{}\xspace}

\def\Kp      {{\ensuremath{\kaon^+}}\xspace}
\def\Km      {{\ensuremath{\kaon^-}}\xspace}

\def\KS      {{\ensuremath{\kaon^0_{\mathrm{S}}}}\xspace}

\def\Kstarz  {{\ensuremath{\kaon^{*0}}}\xspace}
\def\Kstarz892  {{\ensuremath{\kaon^{*}(892)^{0}}}\xspace}
\def\Kstarzb {{\ensuremath{\Kbar{}^{*0}}}\xspace}
\def\Kstarzb892 {{\ensuremath{\Kbar{}^{*}(892)^{0}}}\xspace}

  \def\Dbar    {{\kern 0.2em\overline{\kern -0.2em \PD}{}}\xspace}
\def\D       {{\ensuremath{\PD}}\xspace}

\def\DorDbar {\kern 0.18em\optbar{\kern -0.18em D}{}\xspace}

\def\Dp      {{\ensuremath{\D^+}}\xspace}

\def\B       {{\ensuremath{\PB}}\xspace}
\def\Bbar    {{\ensuremath{\kern 0.18em\overline{\kern -0.18em \PB}{}}}\xspace}

\def\BorBbar    {\kern 0.18em\optbar{\kern -0.18em B}{}\xspace}

\def\Bu      {{\ensuremath{\B^+}}\xspace}

\def\jpsi     {{\ensuremath{{\PJ\mskip -3mu/\mskip -2mu\Ppsi\mskip 2mu}}}\xspace}

\def\Y#1S{\ensuremath{\PUpsilon{(#1S)}}\xspace}

\def\proton      {{\ensuremath{\Pp}}\xspace}

\def\Lz          {{\ensuremath{\PLambda}}\xspace}

\def\LorLbar     {\kern 0.18em\optbar{\kern -0.18em \PLambda}{}\xspace}
\def\Lambdac     {{\ensuremath{\PLambda^+_\cquark}}\xspace}

\def\Xires       {{\ensuremath{\PXi}}\xspace}

\def\Lc          {{\ensuremath{\Lz^+_\cquark}}\xspace}

\def\Xic         {{\ensuremath{\Xires_\cquark}}\xspace}
\def\Xicz        {{\ensuremath{\Xires^0_\cquark}}\xspace}
\def\Xicp        {{\ensuremath{\Xires^+_\cquark}}\xspace}

\def\Xicc        {{\ensuremath{\Xires^{+(+)}_{\cquark\cquark}}}\xspace}

\def\Xiccp       {{\ensuremath{\Xires^+_{\cquark\cquark}}}\xspace}

\def\Xiccpp      {{\ensuremath{\Xires^{++}_{\cquark\cquark}}}\xspace}

\def\Xic2645{{\ensuremath{\Xires_{\cquark}(2645)^0}}\xspace}

\def\BF         {{\ensuremath{\mathcal{B}}}\xspace}

\newcommand{\decay}[2]{\mbox{\ensuremath{#1\!\to #2}}\xspace}         % {\Pa}{\Pb \Pc}

\def\to                 {\ensuremath{\rightarrow}\xspace}

\def\eps   {{\ensuremath{\varepsilon}}\xspace}

\def\XiccpLc    {\decay{\Xiccp}{\Lambdac\Km\pip}}
\def\XiccpDp    {\decay{\Xiccp}{\proton\Dp\Km}}
\def\XiccppLc   {\decay{\Xiccpp}{\Lambdac\Km\pip\pip}}
\def\XiccppDp   {\decay{\Xiccpp}{\Dp\proton\Km\pip}}

\def\XiccpXiczSh  {\decay{\Xiccp}{\Xicz\pip}}
\def\XiccpXicpSh  {\decay{\Xiccp}{\Xicp\pim\pip}}

\def\WSM  {\Xicp\pim\pim}

\def\XiccpXicpRes  {\decay{\Xiccp}{\decay{(\Xic2645}{\Xicp\pim)}\pip}}

\def\XiccpXicpRho {\decay{\Xiccp}{\Xicp\decay{(\rhoz}{\pim\pip)}}}
\def\XiccpXicpRhoSh {\decay{\Xiccp}{\Xicp\rhoz}}

\def\XiccppXicp  {\decay{\Xiccpp}{\decay{(\Xicp}{\proton\Km\pip)}\pip}}
\def\XiccppXicpSh  {\decay{\Xiccpp}{\Xicp\pip}}

\def\XicpDecay {\decay{\Xicp}{\proton\Km\pip}}
\def\DpDecay   {\decay{\Dp}{\Km\pip\pip}}

\def\AT#1     {\ensuremath{A_{\mathrm{T}}^{#1}}\xspace}           % 2

\def\C#1      {\ensuremath{\mathcal{C}_{#1}}\xspace}                       % 9
\def\Cp#1     {\ensuremath{\mathcal{C}_{#1}^{'}}\xspace}                    % 7
\def\Ceff#1   {\ensuremath{\mathcal{C}_{#1}^{\mathrm{(eff)}}}\xspace}        % 9  
\def\Cpeff#1  {\ensuremath{\mathcal{C}_{#1}^{'\mathrm{(eff)}}}\xspace}       % 7
\def\Ope#1    {\ensuremath{\mathcal{O}_{#1}}\xspace}                       % 2
\def\Opep#1   {\ensuremath{\mathcal{O}_{#1}^{'}}\xspace}                    % 7

\newcommand{\aunit}[1]{\ensuremath{\text{\,#1}}}       
\newcommand{\tev}{\aunit{Te\kern -0.1em V}\xspace}
\newcommand{\gev}{\aunit{Ge\kern -0.1em V}\xspace}
\newcommand{\mev}{\aunit{Me\kern -0.1em V}\xspace}
\newcommand{\kev}{\aunit{ke\kern -0.1em V}\xspace}
\newcommand{\ev}{\aunit{e\kern -0.1em V}\xspace}
 
\newcommand{\mevc}{\ensuremath{\aunit{Me\kern -0.1em V\!/}c}\xspace}
\newcommand{\gevc}{\ensuremath{\aunit{Ge\kern -0.1em V\!/}c}\xspace}
\newcommand{\mevcc}{\ensuremath{\aunit{Me\kern -0.1em V\!/}c^2}\xspace}
\newcommand{\gevcc}{\ensuremath{\aunit{Ge\kern -0.1em V\!/}c^2}\xspace}
\newcommand{\mevccinv}{\ensuremath{c^2\!/\aunit{Me\kern -0.1em V}}\xspace}

\def\mum  {\ensuremath{{\,\upmu\mathrm{m}}}\xspace}

\def\invfb   {\ensuremath{\mbox{\,fb}^{-1}}\xspace}

\def\ps   {\ensuremath{{\mathrm{ \,ps}}}\xspace}
\def\fs   {\ensuremath{\mathrm{ \,fs}}\xspace}

\newcommand{\stat}{\ensuremath{\mathrm{\,(stat)}}\xspace}
\newcommand{\syst}{\ensuremath{\mathrm{\,(syst)}}\xspace}

\newcommand{\chisq}{\ensuremath{\chi^2}\xspace}
\newcommand{\chisqndf}{\ensuremath{\chi^2/\mathrm{ndf}}\xspace}
\newcommand{\chisqip}{\ensuremath{\chi^2_{\text{IP}}}\xspace}

\def\gsim{{~\raise.15em\hbox{$>$}\kern-.85em
          \lower.35em\hbox{$\sim$}~}\xspace}
\def\lsim{{~\raise.15em\hbox{$<$}\kern-.85em
          \lower.35em\hbox{$\sim$}~}\xspace}

\def\pt         {\ensuremath{p_{\mathrm{T}}}\xspace}
\def\ptot       {\ensuremath{p}\xspace}

\def\mrad{\ensuremath{\mathrm{ \,mrad}}\xspace}

\def\evtgen     {\mbox{\textsc{EvtGen}}\xspace}

\def\geant      {\mbox{\textsc{Geant4}}\xspace}

\def\photos     {\mbox{\textsc{Photos}}\xspace}

\def\pythia     {\mbox{\textsc{Pythia}}\xspace}
\def\genxicctwo     {\mbox{\textsc{GenXicc2.0}}\xspace}

\def\tell1  {TELL1\xspace}
\def\ukl1   {UKL1\xspace}

\usepackage{cite} % Allows for ranges in citations
\usepackage{mciteplus}
\usepackage{longtable} % only for template; not usually to be used in PAPERs

\begin{document}

%%%%%%%%%%%%%%%%%%%%%%%%%
%%%%% Title     %%%%%%%%%
%%%%%%%%%%%%%%%%%%%%%%%%%
\renewcommand{\thefootnote}{\fnsymbol{footnote}}
\setcounter{footnote}{1}

% %%%%%%% CHOOSE TITLE PAGE--------
%\onecolumn
%\input{title-LHCb-INT}
%\input{title-LHCb-ANA}
%\input{title-LHCb-CONF}
%\input{title-LHCb-FIGURE}
% ===============================================================================
% Purpose: LHCb-PAPER journal paper title page template
% Author: 
% Created on: 2010-09-25
% ===============================================================================

%%%%%%%%%%%%%%%%%%%%%%%%%
%%%%%  TITLE PAGE  %%%%%%
%%%%%%%%%%%%%%%%%%%%%%%%%
\begin{titlepage}
\pagenumbering{roman}

% Header ---------------------------------------------------
\vspace*{-1.5cm}
\centerline{\large EUROPEAN ORGANIZATION FOR NUCLEAR RESEARCH (CERN)}
\vspace*{1.5cm}
\noindent
\begin{tabular*}{\linewidth}{lc@{\extracolsep{\fill}}r@{\extracolsep{0pt}}}
\ifthenelse{\boolean{pdflatex}}% Logo format choice
{\vspace*{-1.5cm}\mbox{\!\!\!\includegraphics[width=.14\textwidth]{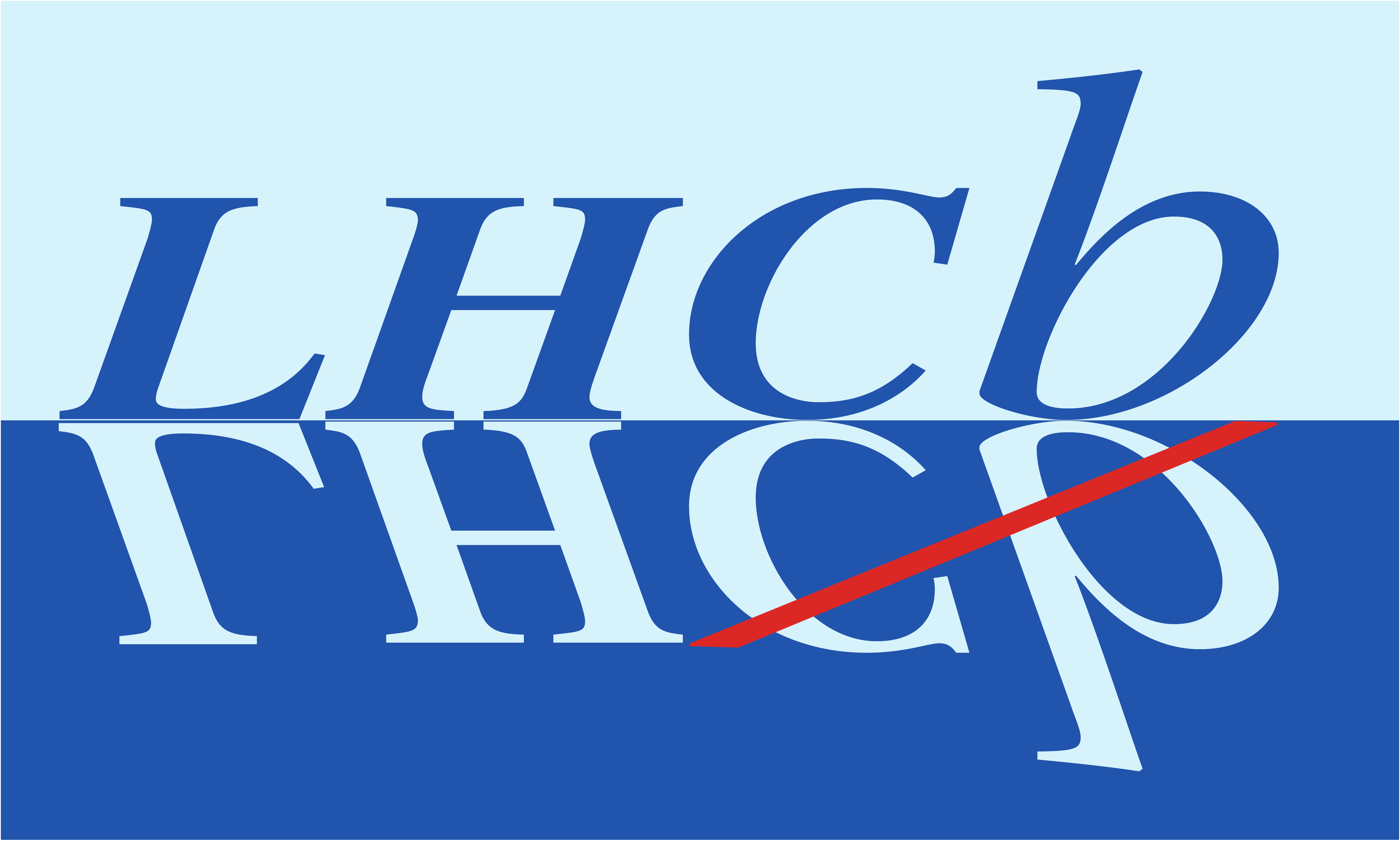}} & &}%
{\vspace*{-1.2cm}\mbox{\!\!\!\includegraphics[width=.12\textwidth]{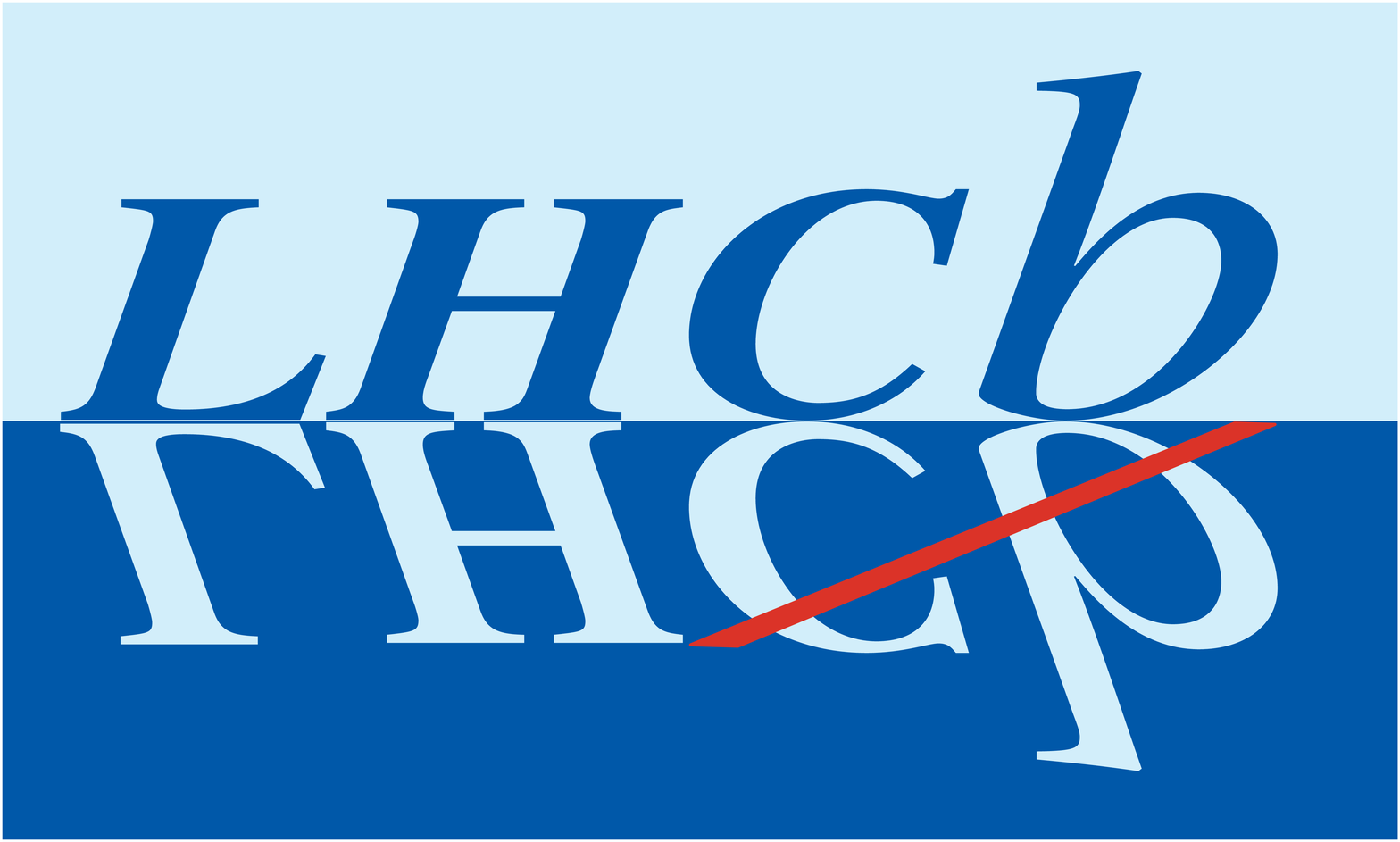}} & &}%
\\
 & & CERN-EP-2021-155 \\  % ID 
 & & LHCb-PAPER-2021-019 \\  % ID 
 & & December 30, 2021 \\ % Date - Can also hardwire e.g.: 23 March 2010
 & & \\
% not in paper \hline
\end{tabular*}

\vspace*{3.0cm}

% Title --------------------------------------------------
{\normalfont\bfseries\boldmath\huge
\begin{center}
% DO NOT EDIT HERE. Instead edit macro in main.tex to keep metadata correct
  \papertitle 
\end{center}
}

\vspace*{2.0cm}

% Authors -------------------------------------------------
\begin{center}
%In the footnote, replace 'paper' by 'Letter' in case of submission to PRL or PLB 
% Edit macro in main.tex to keep metadata correct
\paperauthors\footnote{Authors are listed at the end of this paper.}
\end{center}

\vspace{\fill}

% Abstract -----------------------------------------------
\begin{abstract}
  \noindent
A search for the doubly charmed baryon \Xiccp is performed in the \Xicp\pim\pip invariant-mass spectrum, where the \Xicp baryon is reconstructed in the \proton\Km\pip final state. The study uses proton-proton collision data collected with the \lhcb detector at a centre-of-mass energy of 13\tev, corresponding to a total integrated luminosity of 5.4\invfb. No significant signal is observed in the invariant-mass range of 3.4--3.8\gevcc. Upper limits are set on the ratio of branching fractions multiplied by the production cross-section with respect to the \XiccppXicp decay for different \Xiccp mass and lifetime hypotheses in the rapidity range from 2.0 to 4.5 and the transverse momentum range from 2.5 to 25\gevc.
The results from this search are combined with a previously published search for the \XiccpLc decay mode, yielding a maximum local significance of 4.0 standard deviations around the mass of 3620\mevcc, including systematic uncertainties. Taking into account the look-elsewhere effect in the 3.5--3.7\gevcc mass window, the combined global significance is 2.9 standard deviations including systematic uncertainties.
  
\end{abstract}

\vspace*{2.0cm}

\begin{center}
  Published in JHEP 12 (2021) 107
\end{center}

\vspace{\fill}

{\footnotesize 
% Edit macro in main.tex to keep metadata correct
\centerline{\copyright~\papercopyright. \href{\paperlicenceurl}{\paperlicence}.}}
\vspace*{2mm}

\end{titlepage}

%%%%%%%%%%%%%%%%%%%%%%%%%%%%%%%%
%%%%%  EOD OF TITLE PAGE  %%%%%%
%%%%%%%%%%%%%%%%%%%%%%%%%%%%%%%%

%  empty page follows the title page ----
\newpage
\setcounter{page}{2}
\mbox{~}
%\newpage
%
%% Author List ----------------------------
%%  You need to get a new author list!
%\input{LHCb_authorlist.tex}
%
%The author list for journal publications is provided by the Membership Committee shortly after 'approval to go to paper' has been given.
%%It will be made available on the page
%%\verb!http://www.physik.uzh.ch/~strauman/forMemCo/LHCb-PAPER-XXXX-XXX/! .
%It will be sent to you by email shortly after a paper number has beens assigned.
%The author list should be included already at first circulation, 
%to allow new members of the collaboration to verify whether they have been included correctly.
%Occasionally a misspelled name is corrected or associated institutions become full members.
%In that case, a new author list will be sent to you.
%In case line numbering doesn't work well after including the authorlist, try moving the \verb!\bigskip! after the last author to a separate line.
%
%
%The authorship for Conference Reports should be ``The LHCb
%  collaboration'', with a footnote giving the name(s) of the contact
%  author(s), but without the full list of collaboration names.

%\twocolumn
% %%%%%%%%%%%%% ---------

\renewcommand{\thefootnote}{\arabic{footnote}}
\setcounter{footnote}{0}

%%%%%%%%%%%%%%%%%%%%%%%%%%%%%%%%
%%%%%  Table of Content   %%%%%%
%%%%%%%%%%%%%%%%%%%%%%%%%%%%%%%%
%%%% Uncomment if desired
%\tableofcontents
\cleardoublepage

%%%%%%%%%%%%%%%%%%%%%%%%%
%%%%% Main text %%%%%%%%%
%%%%%%%%%%%%%%%%%%%%%%%%%

\pagestyle{plain} % restore page numbers for the main text
\setcounter{page}{1}
\pagenumbering{arabic}

%% Uncomment during review phase. 
%% Comment before a final submission.
%\linenumbers

%% This is the main body
%% It is useful to have a single file so comemnts are not missed in overleaf.
\section{Introduction}
\label{sec:Introduction}

The quark model~\cite{GellMann:1964nj,Zweig:352337,*Zweig:570209} predicts the existence of multiplets of baryon and meson states with a structure determined by the symmetry properties of the hadron wave functions. Baryons containing two heavy quarks provide a system for unique tests of phenomenological models and calculation techniques in quantum chromodynamics (QCD). 

The first published result on a doubly charmed baryon \Xiccp (quark content \cquark\cquark\dquark) with a mass of $3518.7 \pm 1.7\mevcc$ was reported by the SELEX collaboration in the \XiccpLc and \XiccpDp decay modes~\cite{selex1,selex2}.\footnote{The inclusion of charge-conjugate modes is implied throughout this paper.} However, subsequent searches for the \Xiccp state by the FOCUS~\cite{focus}, \babar~\cite{babar}, and \belle~\cite{belle} experiments showed no evidence for the reported doubly charmed baryon. The \lhcb collaboration performed a search for the \Xiccp baryon in \XiccpLc decays using a data sample corresponding to an integrated luminosity of $0.65\invfb$~\cite{LHCb-PAPER-2013-049}, followed by a recent search using a data sample corresponding to $9\invfb$ of integrated luminosity~\cite{LHCb-PAPER-2019-029}, neither of which yielded any significant signal. 

In 2017 the \lhcb collaboration reported the first observation of the doubly charmed baryon \Xiccpp (quark content \cquark\cquark\uquark) in the \Lambdac\Km\pip\pip invariant-mass spectrum~\cite{LHCb-PAPER-2017-018}. Subsequently, the \Xiccpp baryon was confirmed in the decay mode \XiccppXicpSh~\cite{LHCb-PAPER-2018-026}, whereas no significant signal was observed in the \XiccppDp decay mode~\cite{LHCb-PAPER-2019-011}. Recent \lhcb results on the \Xiccpp baryon include its production measurement~\cite{LHCb-PAPER-2019-035}; lifetime measurement, $0.256^{~+0.024}_{~-0.022} \stat \pm 0.014 \syst \ps$~\cite{LHCb-PAPER-2018-019} consistent with a weak decay; and a precision mass measurement, $3621.55 \pm 0.23 \stat \pm 0.30 \syst \mevcc$~\cite{LHCb-PAPER-2019-037}. Searching for the isospin partner of the already well established \Xiccpp baryon and, more generally, studying doubly heavy baryons are of key importance for completing the baryon spectrum and shedding light on perturbative and non-perturbative QCD dynamics~\cite{potentials}.

Various theoretical calculation techniques, such as lattice QCD~\cite{mass21,lattice,mass20}, models using one light quark and two heavy quarks~\cite{prediction}, QCD sum rules~\cite{mass19,sumrules,mass9,hqet1,mass22}, heavy-quark effective theory~\cite{hqet2}, the bag model~\cite{bagmodel}, or the relativistic quark model~\cite{qm}, have been applied to determine masses of the ground and excited states of the doubly charmed baryons. The majority of theoretical predictions for the masses of the \Xicc ground states are in the range from 3.5 to 3.7\gevcc~\cite{hqet1,hqet2,mass1,mass2,mass3,mass4,mass5,lifetime3,qm,mass6,bagmodel,mass7,mass8,sumrules,mass9,mass10,mass11,mass12,prediction,lattice,mass13,mass14,mass15,mass16,mass17,mass18,mass19,mass20,mass21,mass22}. The mass splitting between the singly and doubly charged \Xicc baryons is predicted to be small, a few \mevcc~\cite{isospinsplit3,isospinsplit,isospinsplit2}, due to isospin symmetry.

Most of the theoretical predictions for the lifetime of the \Xiccp state are in the range from 40 to 250\fs~\cite{lifetime0,lifetime1, lifetime2,lifetime3, lifetime4,lifetime5,lifetime6,mass1,prediction} and have large uncertainties. However, a common feature in most of these theory predictions is that the doubly charged state \Xiccpp is expected to have a lifetime around 2--4 times larger than the singly charged state \Xiccp due to the effect of the destructive Pauli interference of the \cquark-quark decay products and the valence \uquark quark in the initial state. The \Xiccp lifetime is further shortened due to the transition \decay{\cquark\dquark}{\squark\uquark} in the \Xiccp decays, not present in the \Xiccpp decays, which only proceed via the transition \decay{\cquark}{\squark\uquark\dquarkbar}~\cite{lifetime1, lifetime2, lifetime3, lifetime4}. Based on the measured lifetime of the \Xiccpp baryon and the theoretical predictions for the ratio of the \Xiccpp and \Xiccp baryon lifetimes, the expectations for the lifetime of the singly charged state are in the range from 40 to 160\fs. This shorter lifetime makes searches for the \Xiccp baryon more challenging.

\begin{figure}
    \centering
        \includegraphics[width=0.45\textwidth]{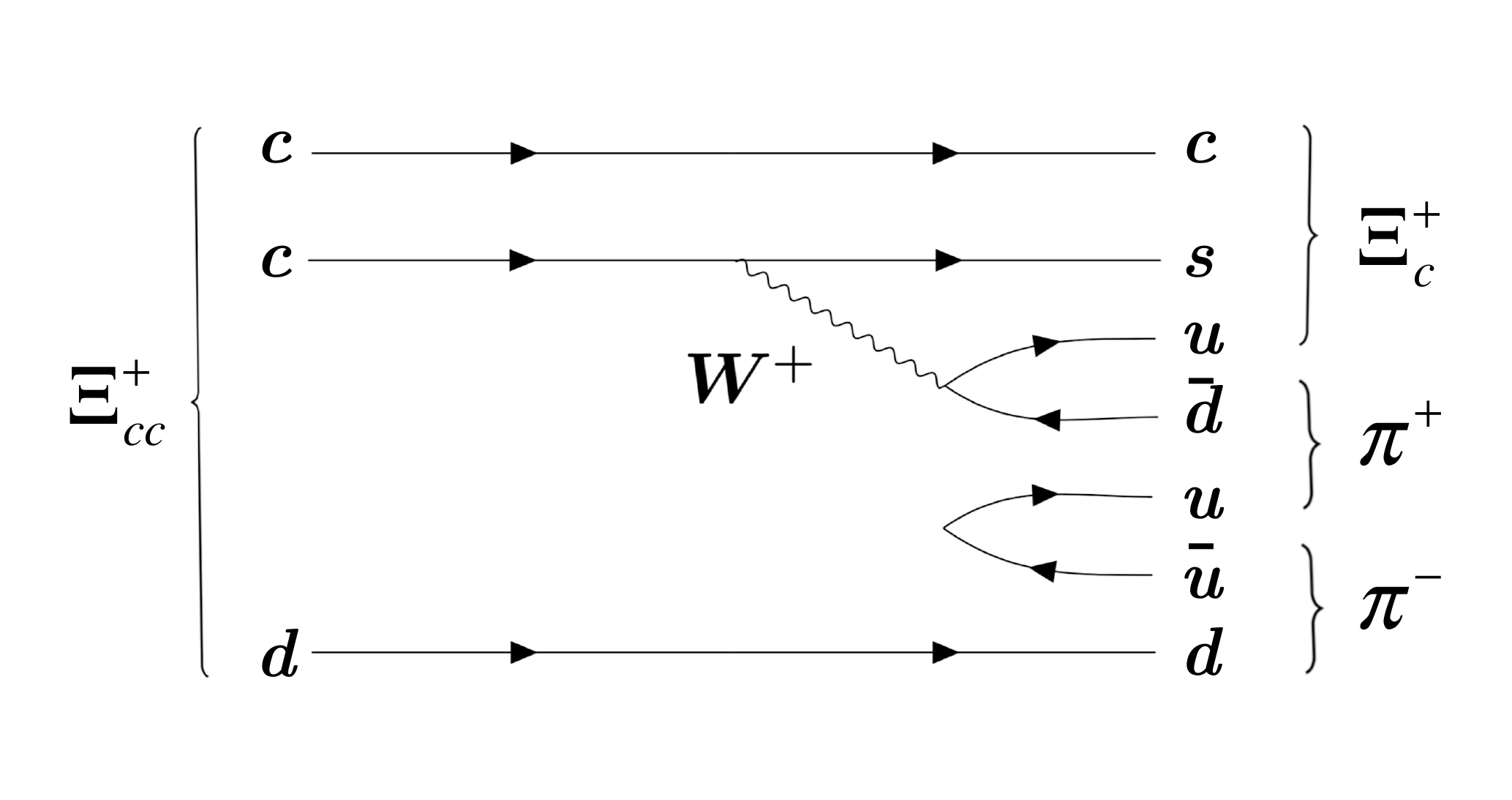}
        \hspace{4mm}
        \includegraphics[width=0.45\textwidth]{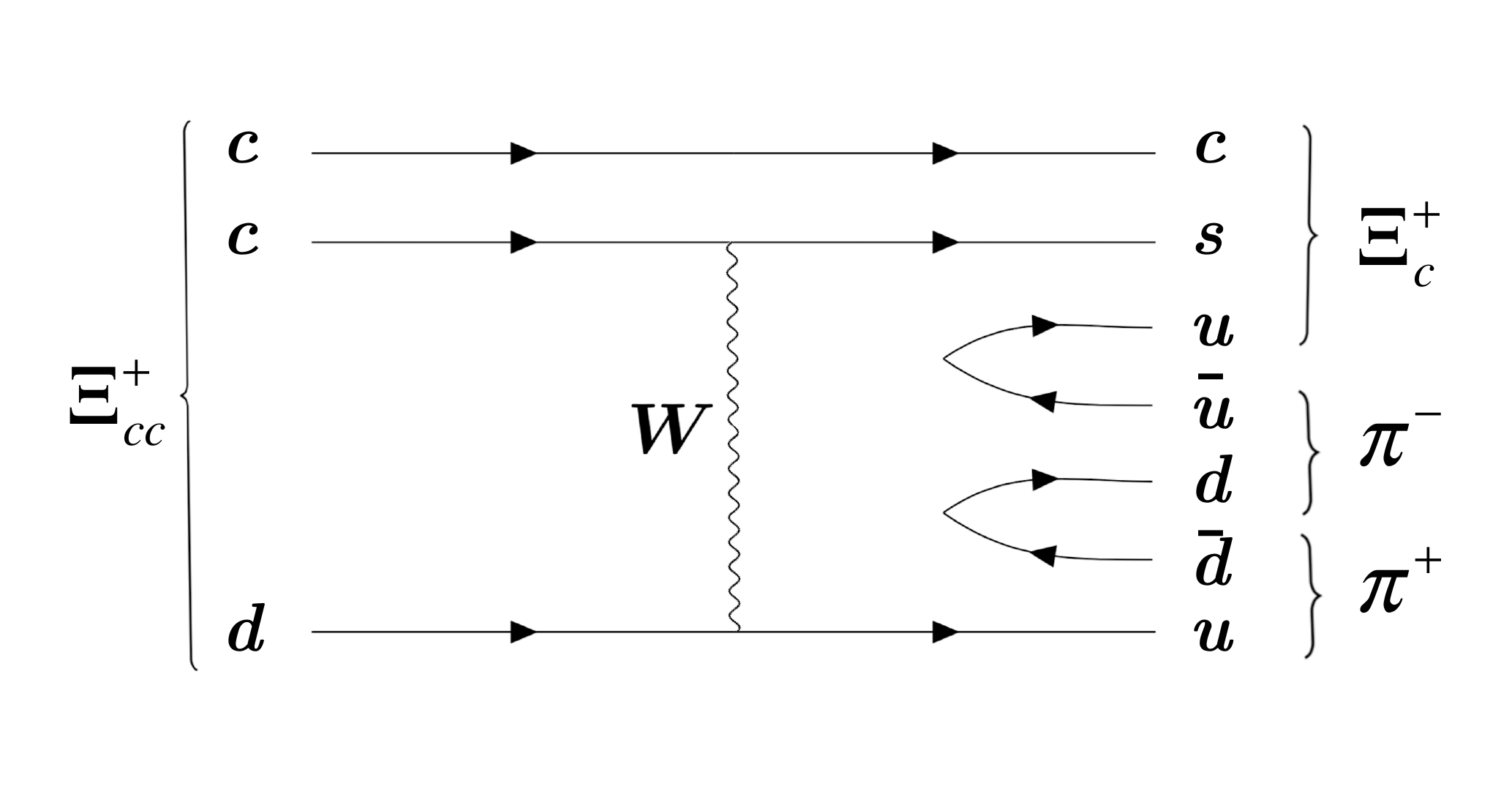}
        \caption{Examples of diagrams for the \XiccpXicpSh decay.}
    \label{fig:feynmanxiccp}
\end{figure}

This paper presents a search for the doubly charmed baryon \Xiccp using the \XiccpXicpSh decay mode where the \Xicp candidates are reconstructed in the \proton\Km\pip final state. The dominant diagrams for this decay are shown in Fig.~\ref{fig:feynmanxiccp}. This decay can  proceed through intermediate resonances, for instance through the \XiccpXicpRho or \XiccpXicpRes decay chains. Since the final state is identical for the studied mode and the two resonant modes, all of these decays are included in the search described in this paper. 
As the branching fraction predictions are commonly calculated for two-body decays, the studied \XiccpXicpSh final state has only an indirect prediction via its resonant decay \XiccpXicpRhoSh, which is indicated as one of the most promising modes to search for the \Xiccp baryon, alongside the \XiccpLc and \XiccpXiczSh decays~\cite{potentials,bftheory2,bftheory3}.
The analysis is based on \proton\proton collision data collected in 2016--2018, corresponding to an integrated luminosity of $5.4\invfb$. 
In order to avoid experimenter's bias, the region of the \Xicp\pim\pip invariant mass from 3.3 to 3.8\gevcc was not examined until the full procedure had been finalised. This range covers both the \Xiccp mass measured by the SELEX experiment and the mass of the \Xiccpp baryon measured by the \lhcb experiment, and most theoretical predictions.

The observed signal yield in the \XiccpXicpSh decay mode is compared to that observed in the already established \XiccppXicpSh channel. This enables a measurement of the ratio of production cross-section times branching fraction between the two channels or setting an upper limit on this quantity. This normalisation mode is chosen to reduce the uncertainty on the ratio of reconstruction and selection efficiencies between the two decays. The production cross-sections of the \Xiccp and \Xiccpp baryons are expected to be the same~\cite{hadronprod}.

The paper is organised as follows. Section~\ref{sec:detector} describes the \lhcb detector and simulation, followed by Sec.~\ref{sec:selection} describing the event selection. Section~\ref{sec:mass} summarises the studies of the mass spectrum, the evaluation of the $p$-values and the combination with the \Xiccp search in the \Lc\Km\pip final state. Section~\ref{sec:limits} describes the determination of the upper limit on the production cross-section multiplied by the branching fraction with respect to the normalisation channel, followed by Sec.~\ref{sec:systematics} with a detailed description of the systematic uncertainties related to the upper limit evaluation. The results are presented and summarised in Secs.~\ref{sec:resultsUL} and~\ref{sec:conclusion}.

\section{Detector and simulation} \label{sec:detector}
The \lhcb detector~\cite{LHCb-DP-2008-001,LHCb-DP-2014-002} is a single-arm forward
spectrometer covering the \mbox{pseudorapidity} range $2<\eta <5$,
designed for the study of particles containing \bquark or \cquark
quarks. The detector includes a high-precision tracking system
consisting of a silicon-strip vertex detector surrounding the $pp$
interaction region~\cite{LHCb-DP-2014-001}, a large-area silicon-strip detector located
upstream of a dipole magnet with a bending power of about
$4{\mathrm{\,Tm}}$, and three stations of silicon-strip detectors and straw
drift tubes~\cite{LHCb-DP-2017-001}
placed downstream of the magnet.
The tracking system provides a measurement of the momentum, \ptot, of charged particles with
a relative uncertainty that varies from 0.5\% at low momentum to 1.0\% at 200\gevc.
The minimum distance of a track to a primary $pp$ collision vertex (PV), the impact parameter (IP), 
is measured with a resolution of $(15+29/\pt)\mum$,
where \pt is the component of the momentum transverse to the beam axis, in\,\gevc.
Different types of charged hadrons are distinguished using information
from two ring-imaging Cherenkov detectors~\cite{LHCb-DP-2012-003}. Photons, electrons and hadrons are identified by a calorimeter system consisting of
scintillating-pad and preshower detectors, an electromagnetic
and a hadronic calorimeter~\cite{LHCb-DP-2020-001}. Muons are identified by a
system composed of alternating layers of iron and multiwire
proportional chambers or triple-GEM detectors~\cite{LHCb-DP-2012-002}.

The online event selection is performed by a trigger~\cite{LHCb-DP-2012-004}, which consists of a hardware stage, followed by a two-level software stage, which applies a full event reconstruction.
In between the two software stages, an alignment and calibration of the detector is performed in near real-time and their results are used in the trigger~\cite{LHCb-PROC-2015-011}.
The same alignment and calibration information is propagated 
to the offline reconstruction, ensuring consistent and high-quality 
particle identification (PID) information between the trigger and 
offline software. The identical performance of the online and offline 
reconstruction offers the opportunity to perform physics analyses 
directly using candidates reconstructed in the trigger 
\cite{LHCb-DP-2012-004,LHCb-DP-2016-001}, which is done in this analysis. 

The momentum of charged particles is calibrated using samples of $\decay{\jpsi}{\mumu}$ 
and $\decay{\Bu}{\jpsi\Kp}$~decays collected concurrently
with the~data sample used for this analysis~\cite{LHCb-PAPER-2012-048,LHCb-PAPER-2013-011}. The~relative accuracy of this procedure is estimated to be $3 \times 10^{-4}$ using samples of other fully reconstructed $\bquark$~hadrons, and $\PUpsilon$~and
$\KS$~mesons.

Simulated \XiccpXicpSh decays are used to optimise the signal selection and to evaluate the efficiencies used for the calculation of the upper limit on the relative production cross-section times branching fraction of the studied decay \XiccpXicpSh compared to the normalisation channel \XiccppXicpSh.  The $pp$ collisions are generated using \pythia~\cite{Sjostrand:2007gs,*Sjostrand:2006za} 
with a specific \lhcb configuration~\cite{LHCb-PROC-2010-056}.
A dedicated generator for doubly heavy baryon production, \genxicctwo~\cite{genxicc}, is used to produce the signal candidates. 
Decays of unstable particles are described by \evtgen~\cite{Lange:2001uf}, in which final-state
radiation is generated using \photos~\cite{davidson2015photos}.
The interaction of the generated particles with the detector, and its response, are implemented using the \geant toolkit~\cite{Allison:2006ve,*Agostinelli:2002hh} as described in Ref.~\cite{LHCb-PROC-2011-006}. The \Xiccp and \Xiccpp baryons are generated with a mass of 3621.4\mevcc. In simulation, the decay products of the \Xiccp and \Xiccpp baryons are distributed uniformly in phase space. The singly charmed \Xicp decays are distributed according to a resonant model in which 55\% of the \Xicp decays proceed via the resonant decay \proton\Kstarzb892 followed by the decay of the \Kstarzb892 meson to \Km\pip final state~\cite{PDG2019}.

\section{Reconstruction and selection} \label{sec:selection}

The event selection is based on four main steps: a trigger selection, an offline selection based on sequential requirements, a multivariate-analysis (MVA) based selection, and a removal of multiple candidates.
The selection is optimised to efficiently retain the \Xiccp signal candidates and to suppress background from random combinations of tracks and from candidates built using misidentified particles.
The optimisation uses simulated events to represent the signal candidates, and the combinatorial background is represented by data with an incorrect combination of charged tracks, the same-sign pions (SSP) \WSM combinations. 
The selection of the \XiccppXicpSh candidates used as normalisation channel is designed to be as similar as possible to the signal channel. 

This analysis uses two different trigger selections: the so-called {\it default trigger set}, which is used for the determination of the upper limit and is applied to both the signal and the normalisation channel in order to reduce the systematic uncertainty on the efficiency ratio between them; and the so-called {\it extended trigger set}, which uses more selected candidates to enhance the probability of observing a significant signal.

The offline candidates can be associated with a hardware trigger decision. The hardware trigger uses information from the muon and calorimeter systems~\cite{LHCb-DP-2019-001}. The events can be selected by the hardware trigger either independently of the reconstructed signal or by the decay products of the signal candidate. The former category is used in the default trigger set, and additional events triggered by the decay products of the singly-charmed \Xicp baryon are accepted in the extended trigger set.

In the software trigger stage of the default trigger set, the \Xicp candidates must be reconstructed and accepted by a dedicated \XicpDecay selection, which selects the \Xicp baryons regardless of whether they are produced in the primary \proton\proton interaction or in a decay at a displaced vertex. 
All tracks from the \Xicp candidates must have \pt larger than 200\mevc, a good track quality, and \chisqip with respect to any PV greater than 6, where \chisqip is defined as the difference in the vertex-fit \chisq of a given PV reconstructed with and without the track or particle under consideration. Additionally, at least one of the three tracks must have $\pt > 1\gevc$ and $\chisqip > 16$ and at least two of the tracks must have $\pt > 400\mevc$ and $\chisqip > 9$. The final state tracks are required to be reliably identified as proton, kaon or pion. Furthermore, the particles identified as protons must have a momentum of at least 10\gevc. The scalar sum of the \pt of the three particles must be larger than 3\gevc. Only the \Xicp candidates with a reconstructed invariant mass in the range of 2392--2543\mevcc, which corresponds to a $\pm 75\mevcc$ window around the known \Xicp mass~\cite{PDG2019}, are retained. The \Xicp candidates must have a good vertex-fit quality and point back to their associated PV, with the angle between the vector from the PV to the decay vertex of the \Xicp baryon and the momentum vector of the \Xicp baryon reconstructed from its decay products less than 10\mrad. The associated PV is the one that best fits the flight direction of the reconstructed candidate. The \Xicp decay vertex must be displaced from the associated PV with a distance corresponding to a decay time of at least 0.15\ps. All candidates are required to pass a MatrixNet classifier~\cite{LHCb-DP-2019-001} within the software trigger, which has been trained to identify particles with large \pt, and a decay vertex with a significant displacement from any PV. The \Xiccp candidates are formed offline from the selected \Xicp candidates combined with two oppositely charged particles identified as pions with momenta larger than 2\gevc, $\pt > 200\mevc$ and a good track quality.

The extended trigger set includes in addition to the selection in the default trigger set two other software trigger selections for a subset of the running periods: one additional selection of the \Xiccp candidates, similar to the default trigger selection; and the selection of the \Xicp candidates using a multivariate algorithm~\cite{BBDT,LHCb-PROC-2015-018} trained to identify the \Xicp candidates originating from any baryon decay.

The first stage of the offline selection consists of a set of sequential requirements applied before the MVA selection. All tracks are required to have momenta between 2 and 150\gevc and be in the pseudorapidity range from 1.5 to 5.0. The \Xiccp candidates must have a good vertex-fit quality and point back to the associated PV.
The reconstructed masses of the \Xicp candidates are required to be in the range of 2450--2488\mevcc, which corresponds to $\pm3$ times the mass resolution around its known mass of 2467.93\mevcc~\cite{PDG2019}. The fiducial region is defined in the same way for both signal and normalisation modes: only the \Xiccp and \Xiccpp candidates in the rapidity range from 2.0 to 4.5 and a \pt from 2.5 to 25\gevc are considered.

After the above requirements, the invariant mass of the final state particles originating from the \Xicp vertex is recalculated under a different mass hypothesis for the SSP data in order to reveal misidentified decays. The most common background of this type is from \DpDecay decays where a pion is misidentified as a proton. These misidentified decays are removed by an explicit veto rejecting all candidates in a mass window of 1850--1890\mevcc around the \Dp mass in the $\Km\pip\pip$ invariant-mass spectrum for both the signal and normalisation modes. The veto removes about 20\% of the  background with a signal efficiency of 95\%.

A multidimensional weighting procedure is used on simulated events to simultaneously correct the distributions of the \Xicc \pt, its $\eta$, and the number of tracks in the event, which are the variables where a disagreement between simulation and data is observed. The assumed $J^P$ for the \Xicc states in simulation is $1/2^+$. The weighting procedure uses a gradient boost algorithm~\cite{Rogozhnikov:2016bdp} trained with simulated events and background-subtracted data~\cite{Pivk:2004ty} for the \XiccppLc and \XiccppXicpSh decay channels. As the \Xiccpp and \Xiccp baryons are isospin partners, they are expected to be produced with similar \pt and $\eta$ spectra. Hence the same weighting procedure is applied to the simulation samples for both the signal and normalisation modes to obtain the correction weights, which are then used in the MVA training.

In order to further suppress combinatorial background and increase the signal purity, the second step of the offline selection is an MVA based selection developed using the TMVA package~\cite{Hocker:2007ht}. The MVA classifier is trained using weighted simulated events as a signal proxy and SSP combinations within the mass region of 3500--3700\mevcc as a background proxy, using the candidates from the default trigger set for both simulation and the SSP data. Due to the large size of the SSP data sample, a randomly selected subset corresponding to 5\% of the available data is used in the training.

The variables used in the MVA selection, ordered according to their discriminating power, are: the scalar sum of the \pt of the pions originating from the \Xiccp candidate; the \chisq per degree of freedom (\chisqndf) from a kinematic fit of the decay chain, with a constraint on the \Xicp mass and a requirement on the \Xiccp candidate to originate from the associated PV~\cite{Hulsbergen:2005pu}; the ratio of the \Xicp transverse momentum and the scalar sum of the \pt of the decay products of the \Xiccp candidate; the maximum distance of closest approach (DOCA) between any pairs of the \Xiccp daughters; the \chisqip of the \Xiccp candidate; the \chisqip of the \Xicp candidate; the ratio of the \Xicp \chisqip and the sum of the \chisqip of the decay products of the \Xiccp candidate; the maximum DOCA between any pairs of the \Xicp daughters; the \Xicp vertex \chisqndf; the angle between the vector from the PV to the decay vertex of the \Xiccp candidate and the momentum vector of the \Xiccp candidate reconstructed from its decay products; the scalar sum of the \pt of the decay products of the \Xicp candidate; the ratio of the \Xicp momentum to the scalar sum of the momenta of the decay products of the \Xicp candidate; the vertex \chisqndf of the \Xiccp candidate; the \chisq of the flight distance of the \Xiccp candidate; and the \chisq of the flight distance of the \Xicp candidate.

The MVA selection is performed using a multilayer perceptron classifier~\cite{mlp} and the requirement on its output variable is optimised using the figure of merit introduced in Ref.~\cite{Punzi:2003bu}, with a target significance of five sigma. A Kolmogorov-Smirnov test is applied on the distributions of the output variables from the training and testing samples to verify that the classifier does not show signs of overtraining. The signal efficiency of the MVA selection with respect to the selection applied before the MVA requirement is about 18\% with a background rejection of about 99.9\%.

Two types of multiple \Xiccp candidates are removed after the MVA selection. First, multiple candidates for which at least one track is a clone of another track from the same candidate are removed by requiring the opening angle between any pair of tracks to be larger than 0.5\mrad. Multiple candidates which have the same set of tracks combined differently, e.g. a \pip from the \Xicp decay swapped with the \pip from the \Xiccp decay, are removed. The fraction of the first (second) type of the multiple candidates is around 3.2\% (1.2\%). Only one randomly chosen candidate per event is retained.

An additional requirement on the \Xicp\pim invariant mass is imposed after the full selection is applied, in order to separately evaluate the statistical significance of the decays that proceed through the \Xic2645 resonance. Only the candidates where the \Xicp\pim invariant mass falls in the window of 2635--2660\mevcc, corresponding to twice the mass resolution around the mass of the \Xic2645 resonance~\cite{PDG2019}, are considered in this selection. This gives an additional suppression of the combinatorial background, which increases the sensitivity to this resonant decay mode.

\section{Mass distributions and signal significance} \label{sec:mass}

\begin{figure}
    \centering            \includegraphics[width=0.55\textwidth]{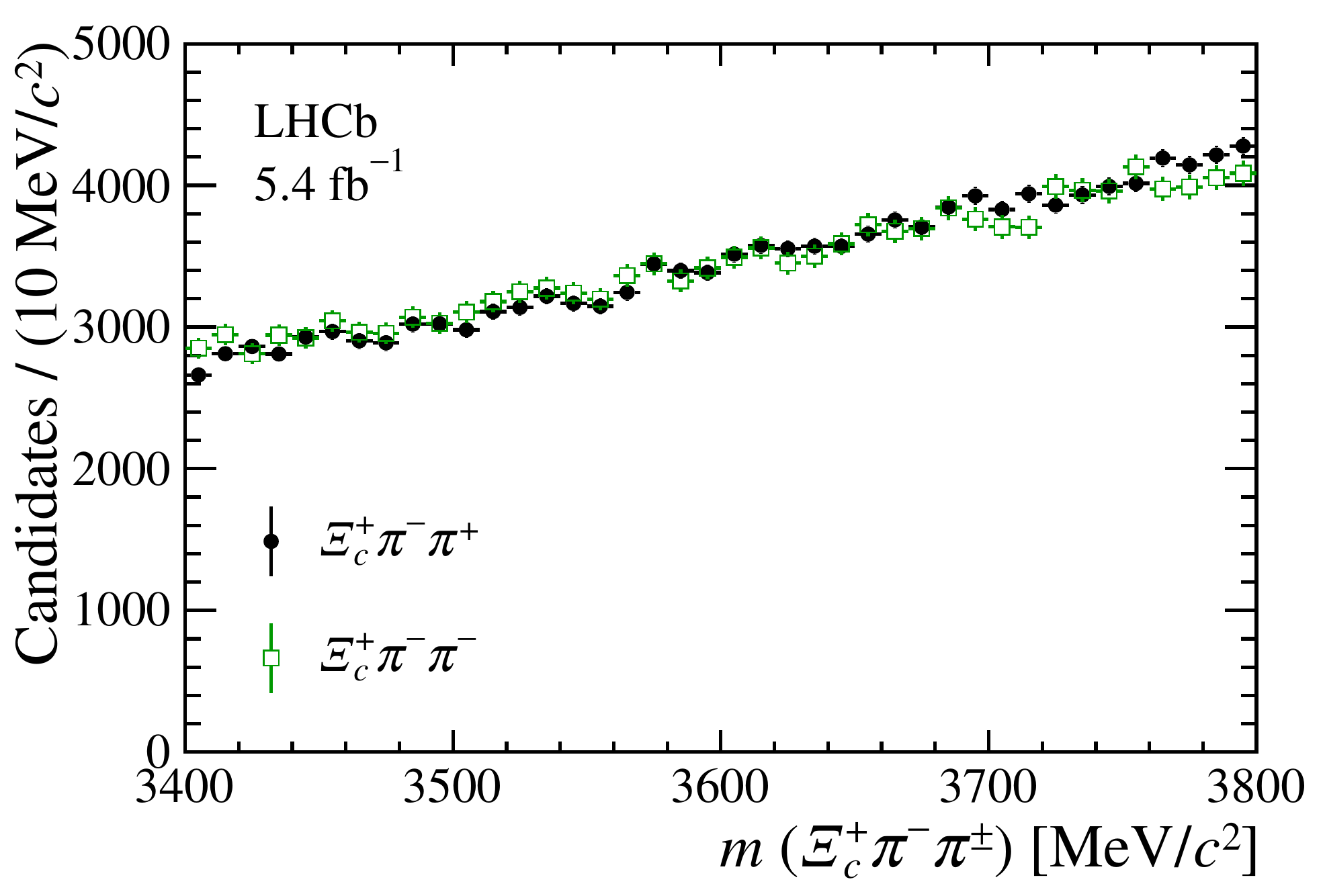}
    \caption{Invariant-mass spectrum for the $\Xicp\pim\pip$ (black points) and the $\Xicp\pim\pim$ (green squares) final states for the default trigger set. The $\Xicp\pim\pim$ data are normalised to the $\Xicp\pim\pip$ invariant-mass spectrum.}
    \label{fig:unblindedfitRSWS}
\end{figure}

\begin{figure}
    \centering            \includegraphics[width=0.55\textwidth]{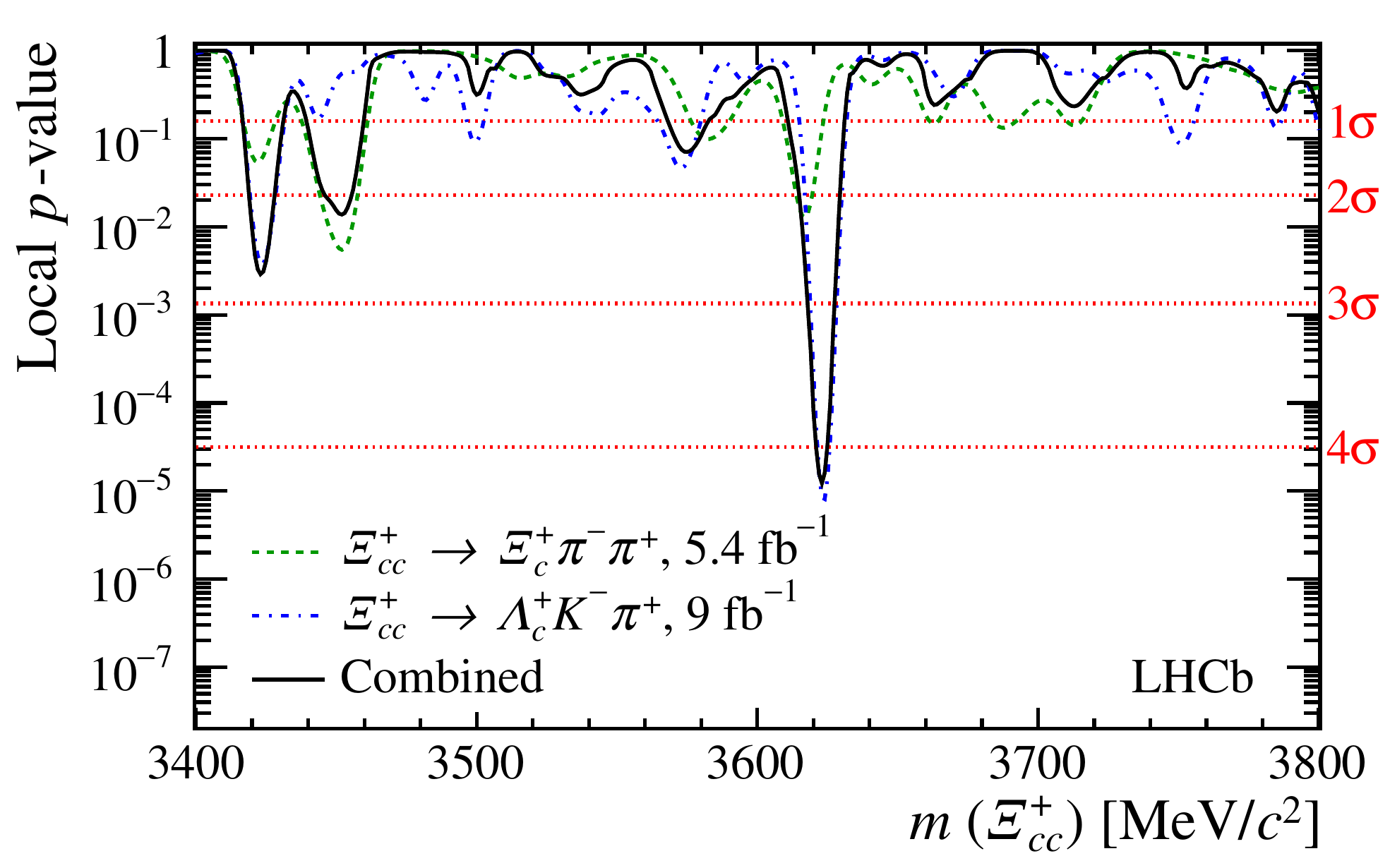}
    \caption{Local $p$-values as a function of the \Xiccp invariant mass, for \Xiccp baryon decays reconstructed in the \Xicp\pim\pip (green dashed curve) and \Lc\Km\pip (blue dash-dotted curve) modes, or combining the two modes (black solid curve). The horizontal dotted red lines represent the $p$-values corresponding to significances of 1, 2, 3 and $4\,\sigma$. The extended trigger set is used for the \XiccpXicpSh decay and Selection B from Ref.~\cite{LHCb-PAPER-2019-029} is used for the \XiccpLc decay. The systematic uncertainties are not taken into account.}
    \label{fig:unblindedpvalue}
\end{figure}

The uncertainty on the mass of the \Xiccp candidate is reduced by measuring the difference in mass between the \Xiccp and \Xicp candidates.
The measured mass of the \Xiccp baryon is given by
\begin{equation}
m (\Xicp\pip\pim) \equiv m([\Xicp\pip\pim]_{\Xiccp})-m([\proton\Km\pip]_{\Xicp})+m(\Xicp),
\label{eq:deltamass}
\end{equation}
where $m([\Xicp\pip\pim]_\Xiccp)$ and $m([\proton\Km\pip]_{\Xicp})$ are the invariant masses of the \Xiccp and \Xicp candidates, and $m(\Xicp)$ is the known mass of the \Xicp baryon~\cite{PDG2019}. 
The distribution of $m(\Xicp\pip\pim)$ after applying the selection and the default trigger set requirements is shown in Fig.~\ref{fig:unblindedfitRSWS}. The $\Xicp\pim\pim$ data are also overlaid for comparison. The local $p$-value is determined as a function of the mass in steps of 1\mevcc.
The $p$-values are determined from the test statistics $q_\pm$, which are based on the ratio of likelihoods of the fit under the background-only and signal-plus-background hypotheses. The test statistics are defined similarly to the test statistic $q_0$ defined in Ref.~\cite{asymptotic}, but contrary to $q_0$ the test statistic $q_\pm$ is assigned the value $-q_0$ when the fit yields a negative number of signal candidates, in order to obtain a smooth $p$-value curve also for downward fluctuations. 
A minimum $p$-value of 0.012, corresponding to a one-sided Gaussian significance of 2.3 standard deviations ($\sigma$), is found at a mass of 3617\mevcc. The $p$-value scan as a function of mass for the extended trigger set is shown in Fig.~\ref{fig:unblindedpvalue}, for which a minimum $p$-value of 0.0024 at a mass of 3452\mevcc, corresponding to $2.8\,\sigma$ local significance, is found. A second minimum is found at the same mass as for the default trigger set, 3617\mevcc, corresponding to a $p$-value of 0.010 and a local significance of $2.3\,\sigma$.
Since no significant signal is observed, the mass spectrum shown in Fig.~\ref{fig:unblindedfitRSWS} is used to evaluate the upper limit on the ratio of branching fractions multiplied by the production cross-section with respect to the \XiccppXicp decay as described in Sec.~\ref{sec:limits}. Additionally, the $p$-value is evaluated when the \Xicp\pim invariant mass is restricted around the mass of the \Xic2645 resonance, in order to search for the  resonant decay \XiccpXicpRes. No evidence for this resonant decay is found.

\begin{figure}
    \centering
        \includegraphics[width=0.45\textwidth]{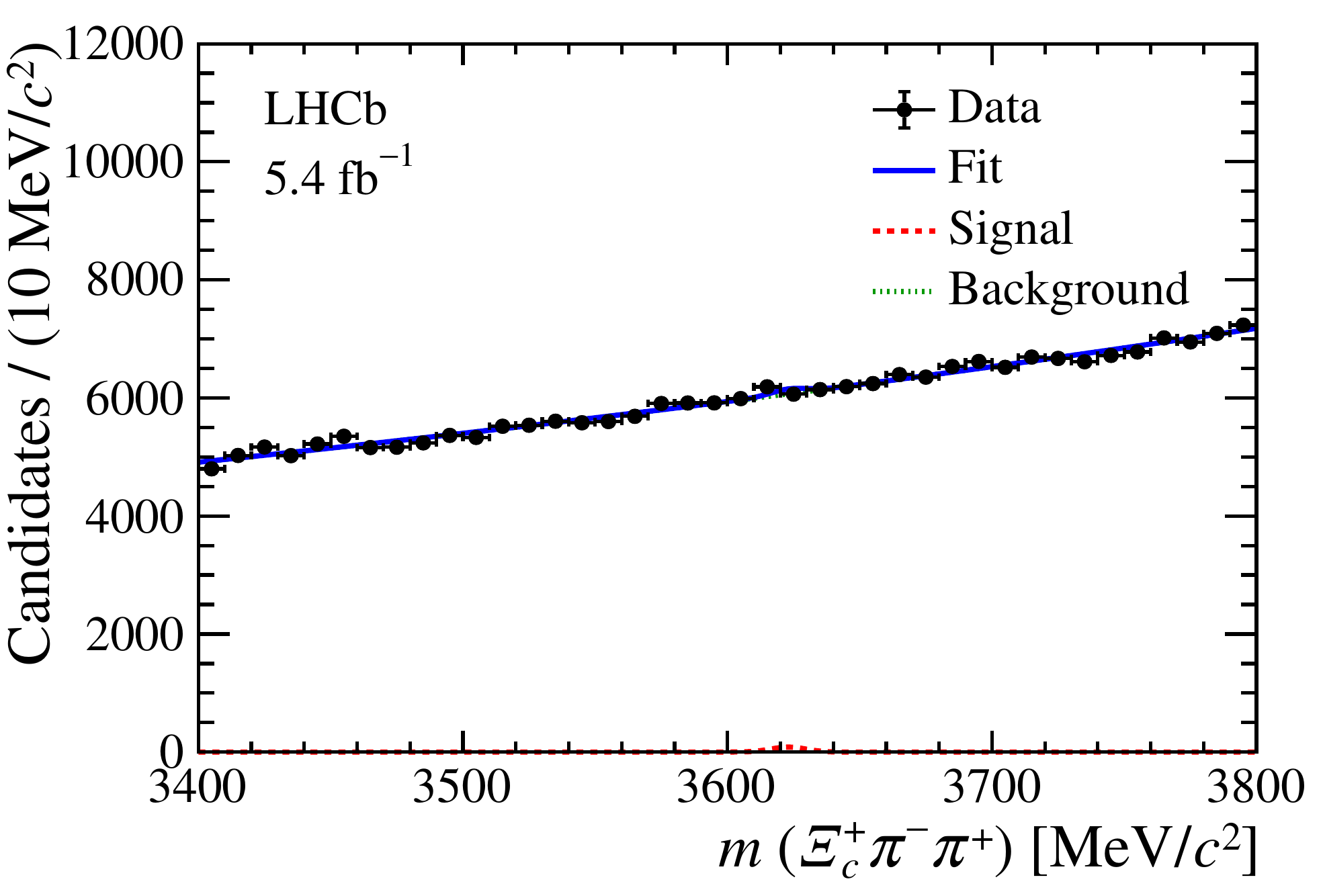}
        \hspace{4mm}
        \includegraphics[width=0.45\textwidth]{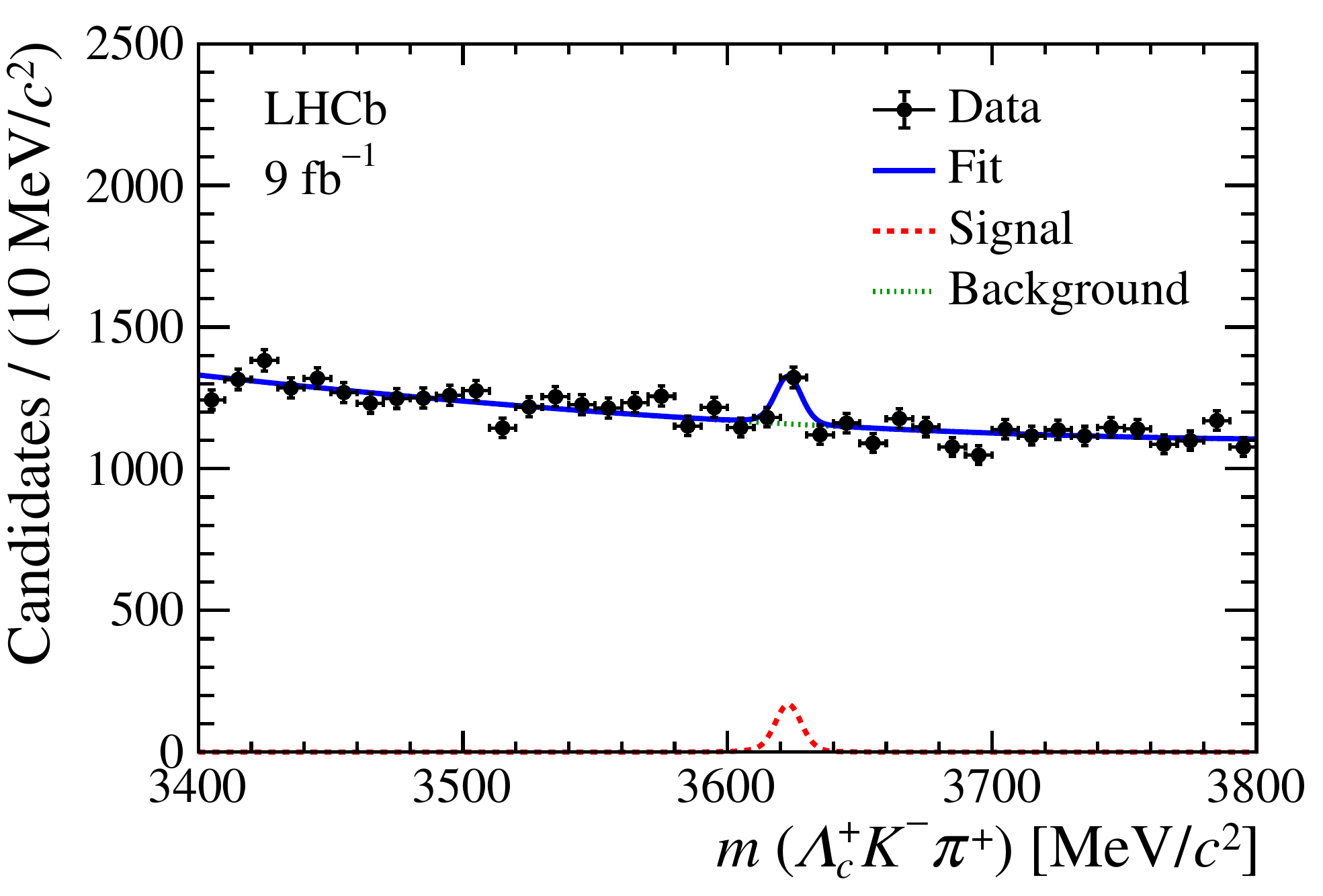}
        \caption{Invariant mass spectra for the (left) \Xicp\pim\pip and (right) \Lambdac\Km\pip final states. The blue solid curve represents the result of a simultaneous fit to the two spectra, with the red dashed (green dotted) curve showing the signal (background) component. The extended trigger set is used for the \XiccpXicpSh decay and Selection B from Ref.~\cite{LHCb-PAPER-2019-029} is used for the \XiccpLc decay.}
    \label{fig:massfit}
\end{figure}

The results from this search are combined with the results from the search for the \Xiccp baryon in the \XiccpLc decay mode presented in Ref.~\cite{LHCb-PAPER-2019-029}. This is performed with a combined fit to the \XiccpXicpSh mass spectrum from the selection with the extended trigger set and the \XiccpLc mass spectrum presented in Ref.~\cite{LHCb-PAPER-2019-029}.\footnote{The sample used here corresponds to the one referred to as Selection B in the referenced paper.} The signal component is modelled with the sum of a Gaussian function and a Crystal Ball function with power-law tails on both sides~\cite{Skwarnicki:1986xj} with a shared mean for both decay channels. An exponential function is used to describe the background contribution for the \XiccpXicpSh decay and a second-order Chebyshev polynomial is used to model the background component for the \XiccpLc decay. The parameters of the signal model are fixed to the values obtained from simulation, all fit parameters in the background model vary freely. Figure~\ref{fig:massfit} shows the \Xicp\pim\pip and \Lambdac\Km\pip invariant-mass spectra, simultaneous unbinned extended maximum-likelihood fit with a common mass and independent signal and background yields is overlaid. The best-fit mass value is $3623.0 \pm 1.4 \mevcc$, where the uncertainty is only statistical, and the signal yield is $223 \pm 54$ for the \XiccpLc decay and $145 \pm 139$ for the \XiccpXicpSh decay.

The combined $p$-values are based on the sum of the test statistics from the two spectra.
Since compatibility with the background-only hypothesis is not unambiguously defined for fits to more than one data set, two alternative methods are used to evaluate the combined $p$-value as cross checks and a good agreement between the methods is found.

The local $p$-values are calculated as a function of mass for both decay channels individually and for the combination, and are shown in Fig.~\ref{fig:unblindedpvalue}. The individual $p$-values are evaluated using the asymptotic formula described in Ref.~\cite{asymptotic} since the distribution of the test statistic $q_\pm$ follows a \chisq distribution with one degree of freedom. The combined $p$-values are calculated from the test statistic in two steps. The $p$-values above $2.2 \times 10^{-4}$ are determined from their corresponding test statistic distributions based on 50\,000 background-only pseudoexperiments. The $p$-values below that value are calculated using the asymptotic formula, which is well described by a $\chi^2$-distribution with two degrees of freedom in that regime.
The minimum $p$-value of $1.2 \times 10^{-5}$, corresponding to a significance of $4.2\,\sigma$ (not including systematic uncertainties), is found at a mass of 3623\mevcc.

Since the mass of the \Xiccp baryon is unknown, the global $p$-value in the 3500--3700\mevcc invariant-mass window is evaluated to account for the look-elsewhere effect. The global $p$-value is computed from 40\,000 background-only pseudoexperiments, and determining what fraction of them has a maximum test statistic larger than the one observed in data. The resulting combined global significance is $3.1\,\sigma$ without accounting for systematic uncertainties.

Three sources of systematic uncertainties are considered when evaluating the combined $p$-values. The first arises from the uncertainty on the relative mass scale between the two mass spectra, which is calculated as the quadratic sum of the uncorrelated systematic uncertainties. They are the uncertainties on the \Lambdac and \Xicp mass and those due to the \Xiccp mass models, resulting in an uncertainty of 0.52\mevcc. The second source of systematic uncertainty is due to a correction on the difference in mass resolution between simulation and data, which is estimated to be 1.37\mevcc for the \XiccpXicpSh decay and 0.70\mevcc for the \XiccpLc decay. 
The last source of uncertainty comes from the choice of fit model, which is evaluated from 10\,000 pseudoexperiments and calculating the difference between the generated yield using an alternative mass model and the fitted yield using the default mass model. This results in a relative uncertainty in the number of signal candidates of 3.1\% for the \XiccpXicpSh decay and 3.3\% for the \XiccpLc decay. The evaluated combined local and global significances including the systematic uncertainties are determined to be $4.0\,\sigma$ and $2.9\,\sigma$, respectively.

\section{Normalisation and single-event sensitivity} \label{sec:limits}

The ratio of production cross-section times the branching fraction between the signal and the normalisation channel is defined as
\begin{equation} \label{eq:bfeff}
R \equiv \frac{\sigma(\Xiccp) \times  \BF (\XiccpXicpSh)}{\sigma(\Xiccpp) \times \BF (\XiccppXicpSh)} = 
\frac{\eps_{\rm{norm}}}{\eps_{\rm{sig}}} \frac{N_{\rm{sig}}}{N_{\rm{norm}}} \equiv	\alpha N_{\rm{sig}},
\end{equation}
where $\sigma(\Xiccp)$ and $\sigma(\Xiccpp)$ are the production cross-sections of the \Xiccp and \Xiccpp baryons, which are expected to be the same~\cite{hadronprod}, and \BF represents the corresponding branching fractions. The number of observed candidates is denoted as $N_{\rm{sig}}$ for the signal channel and $N_{\rm{norm}}$ for the normalisation channel, and the corresponding efficiencies are $\eps_{\rm{sig}}$ and $\eps_{\rm{norm}}$.
The factor $\alpha$ on the right side of Eq.~\ref{eq:bfeff} denotes the single-event sensitivity. 
Since no significant signal is observed for the studied \XiccpXicpSh decay, the upper limit on $R$ is evaluated as a function of assumed \Xiccp mass and for lifetime hypotheses of 40, 80, 120 and 160\fs.

\begin{figure}[t]
    \centering
            \includegraphics[width=0.55\textwidth]{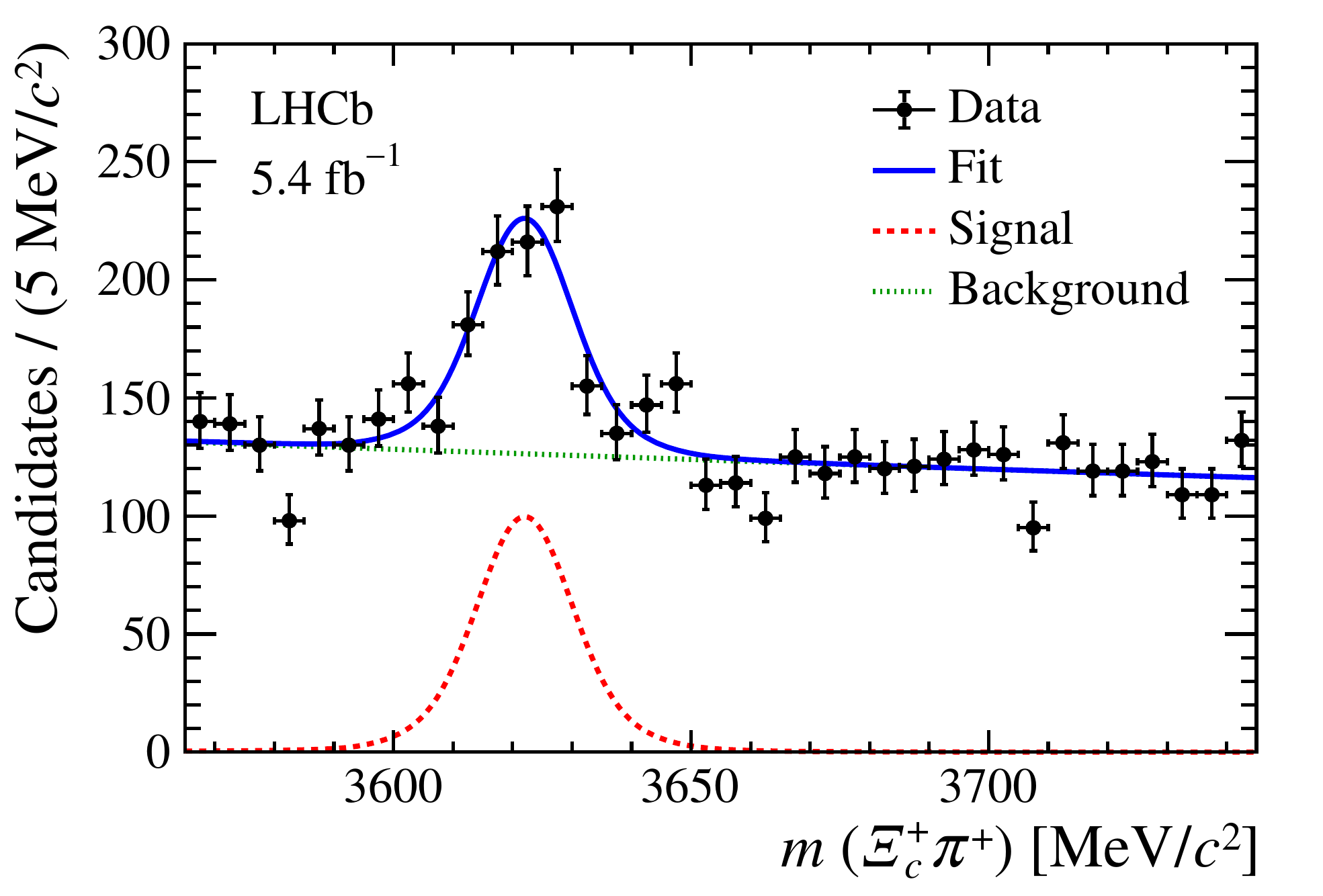}
    \caption{Invariant-mass distribution of the \XiccppXicpSh candidates in the default trigger set. The blue solid line represents the total fit, the red dashed line corresponds to the signal component and the green dotted line to the background component.}
    \label{fig:fitcontroldef}
\end{figure}

There are two main components needed for the evaluation of $\alpha$, the signal yield in the normalisation channel $N_{\rm{norm}}$ and the ratio of efficiencies $\eps_{\rm{norm}} / \eps_{\rm{sig}}$. The invariant-mass distribution of the \Xicp\pip final state is shown in Fig.~\ref{fig:fitcontroldef}, and the signal yield is determined to be $442 \pm 56$ using an extended unbinned maximum-likelihood fit. The signal and background mass fit models are the same as for the signal decay \XiccpXicpSh.

The ratio of efficiencies between the signal and normalisation channels is determined from simulation, where the \Xiccpp baryon lifetime is set to 256\fs, the \Xiccp baryon lifetime is set to 80\fs and both their masses are set to 3621.4\mevcc. Since the lifetime and mass of the \Xiccp baryon are unknown, the variation of the efficiency with the lifetime and mass of the \Xiccp baryon is also considered.
There are three different corrections applied to the overall efficiency ratio: a hardware-trigger correction for possible hardware-trigger mismodelling in simulation; a correction to account for the difference in the fractions of events that are selected by different software trigger categories between simulation and data; and finally a correction due to possible resonant contributions of the \rhoz meson to the \pim\pip spectrum. Moreover, the PID correction, determined in intervals of momentum and pseudorapidity from calibration samples, and tracking correction due to a possible mismodelling of tracking efficiency in the simulation, are evaluated for each individual track.
The resulting corrected efficiency ratio is determined to be $3.63 \pm 0.29$ where the uncertainty is dominated by the total relative systematic uncertainty of 7.9\%, which is described in detail in Sec.~\ref{sec:systematics}. 
The single-event sensitivity $\alpha$ at the lifetime hypothesis of 80\fs, including statistical and systematic uncertainties,  is evaluated to be $0.0082 \pm 0.0012$. Given the unknown lifetime of the \Xiccp baryon, the candidates are weighted to different lifetime hypotheses.
The variation of $\alpha$ with the lifetime, including statistical and systematic uncertainties, is summarised in Table~\ref{tab:alphasfinal}. To determine the variation of the efficiency ratio with the \Xiccp mass, five mass hypotheses are considered in addition to its default mass: 3471, 3521, 3571, 3671 and 3771\mevcc. A linear approximation describes the relation well and is used to determine the efficiency ratio as a function of the assumed \Xiccp mass.

\begin{table}[t]
\caption{Single-event sensitivity including both statistical and systematic uncertainties evaluated for different \Xiccp lifetime hypotheses.} 
\centering 
\begin{tabular}{c|c} 
\toprule[1pt]
Lifetime (\ensuremath{\mathrm{fs}}\xspace) & $\alpha$ \\
\midrule[1pt] 
40 & $0.0122 \pm 0.0018$\\
80 & $0.0082 \pm 0.0012$\\
120 & $0.0067 \pm 0.0010$\\
160 & $0.0061 \pm 0.0009$\\
\bottomrule[1pt] 
\end{tabular}
\label{tab:alphasfinal}
\end{table}

\section{Systematic uncertainties for the upper limits} \label{sec:systematics}

The systematic uncertainties on the ratio of efficiencies are summarised in Table~\ref{tab:effsystem}. All systematic uncertainties on the efficiency ratio are considered to be uncorrelated. Summing them in quadrature gives a total relative systematic uncertainty of 7.9\%. In addition, the systematic uncertainty on the measured signal yield of the normalisation channel is evaluated to be 2.2\% and the systematic uncertainty on the measured background yield for the \XiccpXicpSh decay is found to be 1.3\%.

\begin{table}[t]
\caption{Systematic uncertainties on the ratio of efficiencies for the upper limit determination.}
\centering
\begin{tabular}{l|c} 
\toprule[1pt] 
\multicolumn{1}{l|}{Category} &\multicolumn{1}{c}{Uncertainty on}\\
& $\eps_{\rm{norm}}$/$\eps_{\rm{sig}}$ (\%) \\ 
\midrule[1pt] 
Tracking & 2.7\\
PID &  0.3\\
Hardware trigger & 2.1\\
Simulation correction &  1.7\\
\Xiccpp lifetime & 6.8\\
\XiccpXicpRhoSh contribution & 1.1\\
Limited simulation sample & 1.0\\
\bottomrule[1pt]
Combined & 7.9\\
\bottomrule[1pt]
\end{tabular}
\label{tab:effsystem}
\end{table}

The systematic uncertainty associated with the tracking efficiency comes from the additional pion track in the signal mode that is not present in the normalisation mode, which does not cancel out in the efficiency ratio. It consists of an uncertainty on the \pion meson reconstruction efficiency due to the modelling of hadronic interactions with the detector material~\cite{LHCb-DP-2013-002}, an uncertainty from the correction method itself, and the limited size of the samples used to derive the efficiency correction. All these uncertainties are added in quadrature and the total uncertainty of the tracking efficiency is 2.7\%.

The systematic uncertainty associated with the PID efficiency correction is evaluated by changing the binning scheme in the variables used for this correction, consistently for both signal and normalisation modes. The largest difference in the overall efficiency ratios between different binning schemes was found to be 0.3\%, which is taken as the PID uncertainty.

The hardware-trigger decisions for the signal and normalisation modes are based on information in the event that is independent of their decay products, but there may be a correlation between this information and the kinematic properties of the doubly charmed baryons. The \Xiccp and \Xiccpp baryons are expected to be produced with identical momentum spectra as they are isospin partners so the hardware trigger efficiencies are assumed to be equal for the signal and normalisation modes. However, a difference in kinematics is introduced by the selection, which is corrected for and a systematic uncertainty is evaluated for this correction. The correction is determined by comparing the ratio of the hardware trigger efficiencies in simulation before and after the selection is applied. Half of the correction, 0.9\%, is assigned as a systematic uncertainty. The hardware-trigger efficiency ratio also depends on the \Xiccp lifetime and a 1.8\% systematic uncertainty is assigned to account for the unknown lifetime.

A systematic uncertainty is associated with the imperfect description of the selection variable distributions in simulation and the procedure that is used to correct them. It is evaluated from the difference in efficiency between the three-dimensional weighting in \pt, $\eta$ and number of tracks distributions used in the analysis, and the product of three one-dimensional weightings of the individual variables, resulting in an uncertainty of 1.7\%.

The uncertainty on the measured \Xiccpp lifetime translates into an uncertainty on the efficiency ratio, which is evaluated by weighting the decay-time distribution in simulation to correspond to lifetimes varied by $\pm 1\,\sigma$ around its measured value. The largest variation in the efficiency ratio is 6.8\%, which is assigned as a systematic uncertainty.

Since it is possible that the resonant decay \XiccpXicpRho gives a significant contribution to the final state, the effect of its presence is evaluated as a systematic uncertainty. The \pim\pip invariant-mass spectrum is weighted in simulation to match the \rhoz lineshape, corresponding to the extreme case in which 100\% of the companion pions would come from the \rhoz resonance, and the corresponding efficiency ratio is calculated. A correction for this potential contribution is applied by averaging the efficiency ratio with zero and 100\% resonant contribution. Half of this correction, 1.1\%, is assigned as systematic uncertainty.

The uncertainty on the measured signal yield of the normalisation channel is determined by considering alternative models for the signal and background shapes. A double-Gaussian function is considered as an alternative model for the signal component and a first order Chebyshev polynomial function is considered as an alternative model for the background shape. Pseudoexperiments are generated with the alternative models and fitted with the nominal model and the difference in yields is taken as a systematic uncertainty. The difference is found to be 9.7 when changing the signal model and 0.3 when changing the background model. The normalisation channel yield is thus $N_{\rm{norm}} = 442 \pm 57$ including statistical and systematic uncertainties added in quadrature.

The upper limit determination is also impacted by the background modelling. This uncertainty is estimated to be 1.3\% as the maximum relative difference between the generated and fitted yields across all mass windows, where the generated yields are obtained from pseudoexperiments assuming the alternative background model.

\section{Upper limits} \label{sec:resultsUL}

The upper limits on $R$ in Eq.~\ref{eq:bfeff} are determined using the CLs method~\cite{clsmethod} by using pseudoexperiments. The observed number of signal candidates ($n_{\rm{obs}}$), and the expected number of candidates under the background-only ($n_{\rm{b}}$) and signal-plus-background ($n_{\rm{sb}}$) hypotheses are evaluated in a mass window corresponding to twice the \Xiccp mass resolution of 8.9\mevcc. The single-event sensitivity $\alpha$ is used to relate $n_{\rm{sb}}$ and $R$.

The uncertainty on the single-event sensitivity $\alpha$ is included by sampling values from a Gaussian distribution centred at the value of $\alpha$ with a standard deviation equal to the total uncertainty on $\alpha$.
The systematic uncertainty on the background yield is taken into account by sampling from a Gaussian distribution centred at the number of expected background candidates with a standard deviation corresponding to 1.3\% of the observed background yield. The sampled value is used as the mean of the Poisson distribution used to determine $n_{\rm{b}}$. 
The effects on the upper limit from the uncertainty on the mass resolution are considered by evaluating $n_{\rm{obs}}$, $n_{\rm{b}}$ and $n_{\rm{sb}}$ in mass windows of different widths. The widths are determined by varying the mass resolution within its uncertainty, which is determined to be $\pm1.37\mevcc$ based on the difference in mass resolution between simulation and data for the normalisation mode. The study showed that the larger mass window gives a 13\% larger upper limit, and is used for the evaluation of the upper limit on~$R$.

\begin{figure}
    \centering
            \includegraphics[width=0.55\textwidth] {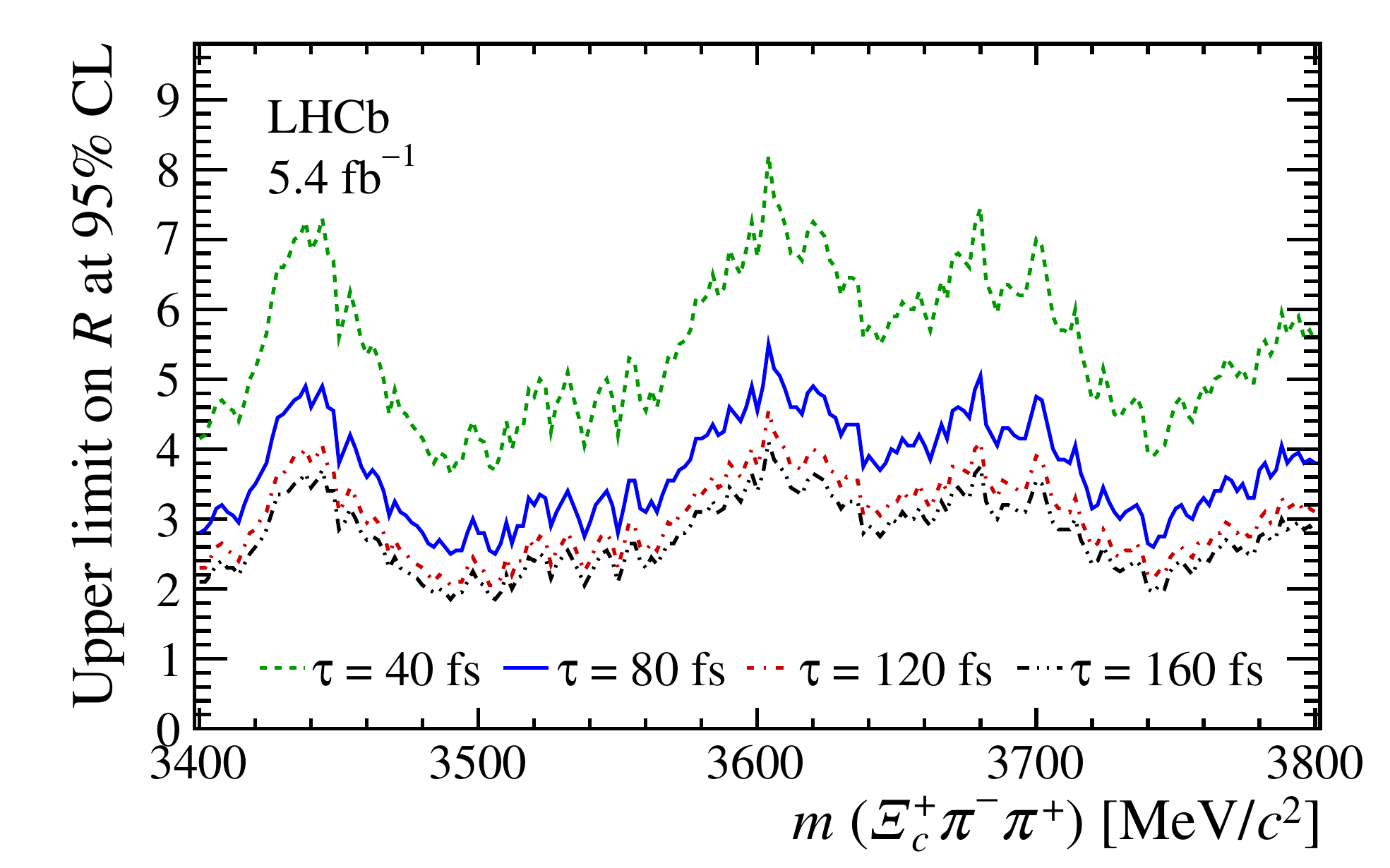}
    \caption{Upper limits on $R$ as a function of the assumed \Xiccp mass for four different lifetime ($\tau$) hypotheses at 95\%~CL.}
    \label{fig:upperlimit95}
\end{figure}

The CLs curve is determined from $3 \times 10^5$ pseudoexperiments for each hypothetical value of $R$ and each mass with 2\mevcc steps in the 3400--3800\mevcc mass window repeated for four lifetime hypotheses. The derived CLs curves are used to determine the upper limits on $R$ at 90 and 95\% confidence levels (CL). 
Figure~\ref{fig:upperlimit95} shows the upper limit on $R$ as a function of mass for the four different lifetime hypotheses at 95\%~CL.

\section{Conclusion} \label{sec:conclusion}

The first search for the \Xiccp baryon in the \Xicp\pim\pip final state is presented and no significant signal is observed. The data used in this search were collected in 2016--2018, corresponding to 5.4\invfb of integrated luminosity. A minimum $p$-value of 0.0108 (0.0024) at a mass of 3617 (3452)\mevcc corresponding to $2.3\,\sigma$ ($2.8\,\sigma$) local significance is found for the default (extended) trigger set in the invariant-mass range of 3400--3800\mevcc.

A combined fit with the \XiccpLc decay mode is performed and the evaluated significances with systematic uncertainties included are $4.0\,\sigma$ for the local and $2.9\,\sigma$ for the global significance. The fitted mass at the minimum $p$-value is $3623.0 \pm 1.4$\mevcc, where the uncertainty is only statistical, for the simultaneous fit to the two spectra, consistent with the mass of the isospin partner \Xiccpp.

Upper limits on $R$, the relative production cross-section times branching fraction of the \XiccpXicpSh decay compared to the normalisation channel \XiccppXicpSh, are determined as a function of assumed \Xiccp masses in the 3400--3800\mevcc mass range for four different lifetime hypotheses. For the mid-range lifetime hypothesis of the \Xiccp baryon of 80\fs, the upper limit on $R$ varies between 2 and 5 at 95\% CL. For the mass with the minimum $p$-value in the combined fit, 3623\mevcc,  the upper limit on $R$ for the \Xiccp lifetime of 80\fs is found to be 4.7 at 95\% CL.
Given the intriguing results presented in this paper, future searches for the \Xiccp baryon in other decay modes using the data already collected by the \lhcb detector are important to clarify the picture. Moreover, a larger data sample will be recorded in the coming years by the upgraded \lhcb detector~\cite{LHCb-TDR-012} which will provide more insight into doubly charmed baryons.

\section*{Acknowledgements}
%
% These Acknowledgements valid from 3-May-2019
%
\noindent We express our gratitude to our colleagues in the CERN
accelerator departments for the excellent performance of the LHC. We
thank the technical and administrative staff at the LHCb
institutes.
We acknowledge support from CERN and from the national agencies:
CAPES, CNPq, FAPERJ and FINEP (Brazil); 
MOST and NSFC (China); 
CNRS/IN2P3 (France); 
BMBF, DFG and MPG (Germany); 
INFN (Italy); 
NWO (Netherlands); 
MNiSW and NCN (Poland); 
MEN/IFA (Romania); 
MSHE (Russia); 
MICINN (Spain); 
SNSF and SER (Switzerland); 
NASU (Ukraine); 
STFC (United Kingdom); 
DOE NP and NSF (USA).
We acknowledge the computing resources that are provided by CERN, IN2P3
(France), KIT and DESY (Germany), INFN (Italy), SURF (Netherlands),
PIC (Spain), GridPP (United Kingdom), RRCKI and Yandex
LLC (Russia), CSCS (Switzerland), IFIN-HH (Romania), CBPF (Brazil),
PL-GRID (Poland) and NERSC (USA).
We are indebted to the communities behind the multiple open-source
software packages on which we depend.
Individual groups or members have received support from
ARC and ARDC (Australia);
AvH Foundation (Germany);
EPLANET, Marie Sk\l{}odowska-Curie Actions and ERC (European Union);
A*MIDEX, ANR, IPhU and Labex P2IO, and R\'{e}gion Auvergne-Rh\^{o}ne-Alpes (France);
Key Research Program of Frontier Sciences of CAS, CAS PIFI, CAS CCEPP, 
Fundamental Research Funds for the Central Universities, 
and Sci. \& Tech. Program of Guangzhou (China);
%Key Research Program of Frontier Sciences of CAS, CAS PIFI,
%Thousand Talents Program, and Sci. \& Tech. Program of Guangzhou (China);
RFBR, RSF and Yandex LLC (Russia);
GVA, XuntaGal and GENCAT (Spain);
the Leverhulme Trust, the Royal Society
 and UKRI (United Kingdom).

% Comment this in for paper drafts; do not include this in analysis note, conference and figure reports
%\input{acknowledgements}

%\input{supplementary}

% This should be taken out in the final paper
%\input{appendix}
%\input{supplementary-app}
%\clearpage

\addcontentsline{toc}{section}{References}
%\setboolean{inbibliography}{true}
\bibliographystyle{LHCb}
\bibliography{main,standard,LHCb-PAPER,LHCb-CONF,LHCb-DP,LHCb-TDR}

\ifx\mcitethebibliography\mciteundefinedmacro
\PackageError{LHCb.bst}{mciteplus.sty has not been loaded}
{This bibstyle requires the use of the mciteplus package.}\fi
\providecommand{\href}[2]{#2}
\begin{mcitethebibliography}{10}
\mciteSetBstSublistMode{n}
\mciteSetBstMaxWidthForm{subitem}{\alph{mcitesubitemcount})}
\mciteSetBstSublistLabelBeginEnd{\mcitemaxwidthsubitemform\space}
{\relax}{\relax}

\bibitem{GellMann:1964nj}
M.~Gell-Mann, \ifthenelse{\boolean{articletitles}}{\emph{{A schematic model of
  baryons and mesons}},
  }{}\href{https://doi.org/10.1016/S0031-9163(64)92001-3}{Phys.\ Lett.\
  \textbf{8} (1964) 214}\relax
\mciteBstWouldAddEndPuncttrue
\mciteSetBstMidEndSepPunct{\mcitedefaultmidpunct}
{\mcitedefaultendpunct}{\mcitedefaultseppunct}\relax
\EndOfBibitem
\bibitem{Zweig:352337}
G.~Zweig, \ifthenelse{\boolean{articletitles}}{\emph{{An SU$_3$ model for
  strong interaction symmetry and its breaking; Version 1}}}{}
  \href{http://cds.cern.ch/record/352337}{CERN-TH-401}, CERN, Geneva,
  1964\relax
\mciteBstWouldAddEndPuncttrue
\mciteSetBstMidEndSepPunct{\mcitedefaultmidpunct}
{\mcitedefaultendpunct}{\mcitedefaultseppunct}\relax
\EndOfBibitem
\bibitem{Zweig:570209}
G.~Zweig, \ifthenelse{\boolean{articletitles}}{\emph{{An SU$_3$ model for
  strong interaction symmetry and its breaking; Version 2}}}{}
  \href{http://cds.cern.ch/record/570209}{CERN-TH-412}, CERN, 1964\relax
\mciteBstWouldAddEndPuncttrue
\mciteSetBstMidEndSepPunct{\mcitedefaultmidpunct}
{\mcitedefaultendpunct}{\mcitedefaultseppunct}\relax
\EndOfBibitem
\bibitem{selex1}
SELEX collaboration, M.~Mattson {\em et~al.},
  \ifthenelse{\boolean{articletitles}}{\emph{{First observation of the doubly
  charmed baryon \Xiccp}},
  }{}\href{https://doi.org/10.1103/PhysRevLett.89.112001}{Phys.\ Rev.\ Lett.\
  \textbf{89} (2002) 112001},
  \href{http://arxiv.org/abs/hep-ex/0208014}{{\normalfont\ttfamily
  arXiv:hep-ex/0208014}}\relax
\mciteBstWouldAddEndPuncttrue
\mciteSetBstMidEndSepPunct{\mcitedefaultmidpunct}
{\mcitedefaultendpunct}{\mcitedefaultseppunct}\relax
\EndOfBibitem
\bibitem{selex2}
SELEX collaboration, A.~Ocherashvili {\em et~al.},
  \ifthenelse{\boolean{articletitles}}{\emph{{Confirmation of the double charm
  baryon \Xiccpselex~via its decay to \proton\Dp\Km}},
  }{}\href{https://doi.org/10.1016/j.physletb.2005.09.043}{Phys.\ Lett.\
  \textbf{B628} (2005) 18},
  \href{http://arxiv.org/abs/hep-ex/0406033}{{\normalfont\ttfamily
  arXiv:hep-ex/0406033}}\relax
\mciteBstWouldAddEndPuncttrue
\mciteSetBstMidEndSepPunct{\mcitedefaultmidpunct}
{\mcitedefaultendpunct}{\mcitedefaultseppunct}\relax
\EndOfBibitem
\bibitem{focus}
S.~P. Ratti {\em et~al.}, \ifthenelse{\boolean{articletitles}}{\emph{{New
  results on c-baryons and a search for cc-baryons in FOCUS}},
  }{}\href{https://doi.org/10.1016/S0920-5632(02)01948-5}{Nucl.\ Phys.\ Proc.\
  Suppl.\  \textbf{115} (2003) 33}\relax
\mciteBstWouldAddEndPuncttrue
\mciteSetBstMidEndSepPunct{\mcitedefaultmidpunct}
{\mcitedefaultendpunct}{\mcitedefaultseppunct}\relax
\EndOfBibitem
\bibitem{babar}
BaBar collaboration, B.~Aubert {\em et~al.},
  \ifthenelse{\boolean{articletitles}}{\emph{{Search for doubly charmed baryons
  \Xiccp and \Xiccpp in BABAR}},
  }{}\href{https://doi.org/10.1103/PhysRevD.74.011103}{Phys.\ Rev.\
  \textbf{D74} (2006) 011103},
  \href{http://arxiv.org/abs/hep-ex/0605075}{{\normalfont\ttfamily
  arXiv:hep-ex/0605075}}\relax
\mciteBstWouldAddEndPuncttrue
\mciteSetBstMidEndSepPunct{\mcitedefaultmidpunct}
{\mcitedefaultendpunct}{\mcitedefaultseppunct}\relax
\EndOfBibitem
\bibitem{belle}
Belle collaboration, R.~Chistov {\em et~al.},
  \ifthenelse{\boolean{articletitles}}{\emph{{Observation of new states
  decaying into \Lambdac\Km\pip and \Lambdac\KS\pim}},
  }{}\href{https://doi.org/10.1103/PhysRevLett.97.162001}{Phys.\ Rev.\ Lett.\
  \textbf{97} (2006) 162001},
  \href{http://arxiv.org/abs/hep-ex/0606051}{{\normalfont\ttfamily
  arXiv:hep-ex/0606051}}\relax
\mciteBstWouldAddEndPuncttrue
\mciteSetBstMidEndSepPunct{\mcitedefaultmidpunct}
{\mcitedefaultendpunct}{\mcitedefaultseppunct}\relax
\EndOfBibitem
\bibitem{LHCb-PAPER-2013-049}
LHCb collaboration, R.~Aaij {\em et~al.},
  \ifthenelse{\boolean{articletitles}}{\emph{{Search for the doubly charmed
  baryon $\Xires_{cc}^+$}},
  }{}\href{https://doi.org/10.1007/JHEP12(2013)090}{JHEP \textbf{12} (2013)
  090}, \href{http://arxiv.org/abs/1310.2538}{{\normalfont\ttfamily
  arXiv:1310.2538}}\relax
\mciteBstWouldAddEndPuncttrue
\mciteSetBstMidEndSepPunct{\mcitedefaultmidpunct}
{\mcitedefaultendpunct}{\mcitedefaultseppunct}\relax
\EndOfBibitem
\bibitem{LHCb-PAPER-2019-029}
LHCb collaboration, R.~Aaij {\em et~al.},
  \ifthenelse{\boolean{articletitles}}{\emph{{Search for the doubly charmed
  baryon $\Xiccp$}}, }{}\href{https://doi.org/10.1007/s11433-019-1471-8}{Sci.\
  China Phys.\ Mech.\ Astron.\  \textbf{63} (2020) 221062},
  \href{http://arxiv.org/abs/1909.12273}{{\normalfont\ttfamily
  arXiv:1909.12273}}\relax
\mciteBstWouldAddEndPuncttrue
\mciteSetBstMidEndSepPunct{\mcitedefaultmidpunct}
{\mcitedefaultendpunct}{\mcitedefaultseppunct}\relax
\EndOfBibitem
\bibitem{LHCb-PAPER-2017-018}
LHCb collaboration, R.~Aaij {\em et~al.},
  \ifthenelse{\boolean{articletitles}}{\emph{{Observation of the doubly charmed
  baryon \Xiccpp}},
  }{}\href{https://doi.org/10.1103/PhysRevLett.119.112001}{Phys.\ Rev.\ Lett.\
  \textbf{119} (2017) 112001},
  \href{http://arxiv.org/abs/1707.01621}{{\normalfont\ttfamily
  arXiv:1707.01621}}\relax
\mciteBstWouldAddEndPuncttrue
\mciteSetBstMidEndSepPunct{\mcitedefaultmidpunct}
{\mcitedefaultendpunct}{\mcitedefaultseppunct}\relax
\EndOfBibitem
\bibitem{LHCb-PAPER-2018-026}
LHCb collaboration, R.~Aaij {\em et~al.},
  \ifthenelse{\boolean{articletitles}}{\emph{{First observation of the doubly
  charmed baryon decay \mbox{\decay{\Xiccpp}{\Xires_c^+\pip}}}},
  }{}\href{https://doi.org/10.1103/PhysRevLett.121.162002}{Phys.\ Rev.\ Lett.\
  \textbf{121} (2018) 162002},
  \href{http://arxiv.org/abs/1807.01919}{{\normalfont\ttfamily
  arXiv:1807.01919}}\relax
\mciteBstWouldAddEndPuncttrue
\mciteSetBstMidEndSepPunct{\mcitedefaultmidpunct}
{\mcitedefaultendpunct}{\mcitedefaultseppunct}\relax
\EndOfBibitem
\bibitem{LHCb-PAPER-2019-011}
LHCb collaboration, R.~Aaij {\em et~al.},
  \ifthenelse{\boolean{articletitles}}{\emph{{A search for
  \mbox{\decay{\Xiccpp}{\Dp p\Km\pip}} decays}},
  }{}\href{https://doi.org/10.1007/JHEP10(2019)124}{JHEP \textbf{10} (2019)
  124}, \href{http://arxiv.org/abs/1905.02421}{{\normalfont\ttfamily
  arXiv:1905.02421}}\relax
\mciteBstWouldAddEndPuncttrue
\mciteSetBstMidEndSepPunct{\mcitedefaultmidpunct}
{\mcitedefaultendpunct}{\mcitedefaultseppunct}\relax
\EndOfBibitem
\bibitem{LHCb-PAPER-2019-035}
LHCb collaboration, R.~Aaij {\em et~al.},
  \ifthenelse{\boolean{articletitles}}{\emph{{Measurement of \Xiccpp production
  in $pp$ collisions at $\sqrt{s} = 13\tev$}},
  }{}\href{https://doi.org/10.1088/1674-1137/44/2/022001}{Chin.\ Phys.\
  \textbf{C44} (2020) 022001},
  \href{http://arxiv.org/abs/1910.11316}{{\normalfont\ttfamily
  arXiv:1910.11316}}\relax
\mciteBstWouldAddEndPuncttrue
\mciteSetBstMidEndSepPunct{\mcitedefaultmidpunct}
{\mcitedefaultendpunct}{\mcitedefaultseppunct}\relax
\EndOfBibitem
\bibitem{LHCb-PAPER-2018-019}
LHCb collaboration, R.~Aaij {\em et~al.},
  \ifthenelse{\boolean{articletitles}}{\emph{{Measurement of the lifetime of
  the doubly charmed baryon \Xiccpp}},
  }{}\href{https://doi.org/10.1103/PhysRevLett.121.052002}{Phys.\ Rev.\ Lett.\
  \textbf{121} (2018) 052002},
  \href{http://arxiv.org/abs/1806.02744}{{\normalfont\ttfamily
  arXiv:1806.02744}}\relax
\mciteBstWouldAddEndPuncttrue
\mciteSetBstMidEndSepPunct{\mcitedefaultmidpunct}
{\mcitedefaultendpunct}{\mcitedefaultseppunct}\relax
\EndOfBibitem
\bibitem{LHCb-PAPER-2019-037}
LHCb collaboration, R.~Aaij {\em et~al.},
  \ifthenelse{\boolean{articletitles}}{\emph{{Precision measurement of the
  $\Xiccpp$ mass}}, }{}\href{https://doi.org/10.1007/JHEP02(2020)049}{JHEP
  \textbf{02} (2020) 049},
  \href{http://arxiv.org/abs/1911.08594}{{\normalfont\ttfamily
  arXiv:1911.08594}}\relax
\mciteBstWouldAddEndPuncttrue
\mciteSetBstMidEndSepPunct{\mcitedefaultmidpunct}
{\mcitedefaultendpunct}{\mcitedefaultseppunct}\relax
\EndOfBibitem
\bibitem{potentials}
F.~Yu {\em et~al.}, \ifthenelse{\boolean{articletitles}}{\emph{{Discovery
  potentials of doubly charmed baryons}},
  }{}\href{https://doi.org/10.1088/1674-1137/42/5/051001}{Chin.\ Phys.\
  \textbf{C42} (2018) 051001},
  \href{http://arxiv.org/abs/1703.09086}{{\normalfont\ttfamily
  arXiv:1703.09086}}\relax
\mciteBstWouldAddEndPuncttrue
\mciteSetBstMidEndSepPunct{\mcitedefaultmidpunct}
{\mcitedefaultendpunct}{\mcitedefaultseppunct}\relax
\EndOfBibitem
\bibitem{mass21}
R.~A. Briceno, H.-W. Lin, and D.~R. Bolton,
  \ifthenelse{\boolean{articletitles}}{\emph{{Charmed-baryon spectroscopy from
  lattice QCD with $N_f=2+1+1$ flavors}},
  }{}\href{https://doi.org/10.1103/PhysRevD.86.094504}{Phys.\ Rev.\
  \textbf{D86} (2012) 094504},
  \href{http://arxiv.org/abs/1207.3536}{{\normalfont\ttfamily
  arXiv:1207.3536}}\relax
\mciteBstWouldAddEndPuncttrue
\mciteSetBstMidEndSepPunct{\mcitedefaultmidpunct}
{\mcitedefaultendpunct}{\mcitedefaultseppunct}\relax
\EndOfBibitem
\bibitem{lattice}
Z.~S. Brown, W.~Detmold, S.~Meinel, and K.~Orginos,
  \ifthenelse{\boolean{articletitles}}{\emph{{Charmed bottom baryon
  spectroscopy from lattice QCD}},
  }{}\href{https://doi.org/10.1103/PhysRevD.90.094507}{Phys.\ Rev.\
  \textbf{D90} (2014) 094507},
  \href{http://arxiv.org/abs/1409.0497}{{\normalfont\ttfamily
  arXiv:1409.0497}}\relax
\mciteBstWouldAddEndPuncttrue
\mciteSetBstMidEndSepPunct{\mcitedefaultmidpunct}
{\mcitedefaultendpunct}{\mcitedefaultseppunct}\relax
\EndOfBibitem
\bibitem{mass20}
C.~Alexandrou {\em et~al.}, \ifthenelse{\boolean{articletitles}}{\emph{{Baryon
  spectrum with $N_f=2+1+1$ twisted mass fermions}},
  }{}\href{https://doi.org/10.1103/PhysRevD.90.074501}{Phys.\ Rev.\
  \textbf{D90} (2014) 074501},
  \href{http://arxiv.org/abs/1406.4310}{{\normalfont\ttfamily
  arXiv:1406.4310}}\relax
\mciteBstWouldAddEndPuncttrue
\mciteSetBstMidEndSepPunct{\mcitedefaultmidpunct}
{\mcitedefaultendpunct}{\mcitedefaultseppunct}\relax
\EndOfBibitem
\bibitem{prediction}
M.~Karliner and J.~L. Rosner,
  \ifthenelse{\boolean{articletitles}}{\emph{{Baryons with two heavy quarks:
  Masses, production, decays, and detection}},
  }{}\href{https://doi.org/10.1103/PhysRevD.90.094007}{Phys.\ Rev.\
  \textbf{D90} (2014) 094007},
  \href{http://arxiv.org/abs/1408.5877}{{\normalfont\ttfamily
  arXiv:1408.5877}}\relax
\mciteBstWouldAddEndPuncttrue
\mciteSetBstMidEndSepPunct{\mcitedefaultmidpunct}
{\mcitedefaultendpunct}{\mcitedefaultseppunct}\relax
\EndOfBibitem
\bibitem{mass19}
D.~B. Lichtenberg, R.~Roncaglia, and E.~Predazzi,
  \ifthenelse{\boolean{articletitles}}{\emph{{Mass sum rules for singly and
  doubly heavy flavored hadrons}},
  }{}\href{https://doi.org/10.1103/PhysRevD.53.6678}{Phys.\ Rev.\  \textbf{D53}
  (1996) 6678},
  \href{http://arxiv.org/abs/hep-ph/9511461}{{\normalfont\ttfamily
  arXiv:hep-ph/9511461}}\relax
\mciteBstWouldAddEndPuncttrue
\mciteSetBstMidEndSepPunct{\mcitedefaultmidpunct}
{\mcitedefaultendpunct}{\mcitedefaultseppunct}\relax
\EndOfBibitem
\bibitem{sumrules}
J.~Zhang and M.~Huang, \ifthenelse{\boolean{articletitles}}{\emph{{Doubly heavy
  baryons in QCD sum rules}},
  }{}\href{https://doi.org/10.1103/PhysRevD.78.094007}{Phys.\ Rev.\
  \textbf{D78} (2008) 094007},
  \href{http://arxiv.org/abs/0810.5396}{{\normalfont\ttfamily
  arXiv:0810.5396}}\relax
\mciteBstWouldAddEndPuncttrue
\mciteSetBstMidEndSepPunct{\mcitedefaultmidpunct}
{\mcitedefaultendpunct}{\mcitedefaultseppunct}\relax
\EndOfBibitem
\bibitem{mass9}
Z.-G. Wang, \ifthenelse{\boolean{articletitles}}{\emph{{Analysis of the
  ${1\over 2}^+$ doubly heavy baryon states with QCD sum rules}},
  }{}\href{https://doi.org/10.1140/epja/i2010-11004-3}{Eur.\ Phys.\ J.\
  \textbf{A45} (2010) 267},
  \href{http://arxiv.org/abs/1001.4693}{{\normalfont\ttfamily
  arXiv:1001.4693}}\relax
\mciteBstWouldAddEndPuncttrue
\mciteSetBstMidEndSepPunct{\mcitedefaultmidpunct}
{\mcitedefaultendpunct}{\mcitedefaultseppunct}\relax
\EndOfBibitem
\bibitem{hqet1}
H.~Chen {\em et~al.}, \ifthenelse{\boolean{articletitles}}{\emph{{Establishing
  low-lying doubly charmed baryons}},
  }{}\href{https://doi.org/10.1103/PhysRevD.96.031501,
  10.1103/PhysRevD.96.119902}{Phys.\ Rev.\  \textbf{D96} (2017) 031501},
  \href{http://arxiv.org/abs/1707.01779}{{\normalfont\ttfamily
  arXiv:1707.01779}}\relax
\mciteBstWouldAddEndPuncttrue
\mciteSetBstMidEndSepPunct{\mcitedefaultmidpunct}
{\mcitedefaultendpunct}{\mcitedefaultseppunct}\relax
\EndOfBibitem
\bibitem{mass22}
Z.-G. Wang, \ifthenelse{\boolean{articletitles}}{\emph{{Analysis of the doubly
  heavy baryon states and pentaquark states with QCD sum rules}},
  }{}\href{https://doi.org/10.1140/epjc/s10052-018-6300-4}{Eur.\ Phys.\ J.\
  \textbf{C78} (2018) 826},
  \href{http://arxiv.org/abs/1808.09820}{{\normalfont\ttfamily
  arXiv:1808.09820}}\relax
\mciteBstWouldAddEndPuncttrue
\mciteSetBstMidEndSepPunct{\mcitedefaultmidpunct}
{\mcitedefaultendpunct}{\mcitedefaultseppunct}\relax
\EndOfBibitem
\bibitem{hqet2}
J.~G. Korner, M.~Kramer, and D.~Pirjol,
  \ifthenelse{\boolean{articletitles}}{\emph{{Heavy baryons}},
  }{}\href{https://doi.org/10.1016/0146-6410(94)90053-1}{Prog.\ Part.\ Nucl.\
  Phys.\  \textbf{33} (1994) 787},
  \href{http://arxiv.org/abs/hep-ph/9406359}{{\normalfont\ttfamily
  arXiv:hep-ph/9406359}}\relax
\mciteBstWouldAddEndPuncttrue
\mciteSetBstMidEndSepPunct{\mcitedefaultmidpunct}
{\mcitedefaultendpunct}{\mcitedefaultseppunct}\relax
\EndOfBibitem
\bibitem{bagmodel}
D.~He {\em et~al.}, \ifthenelse{\boolean{articletitles}}{\emph{{Evaluation of
  spectra of baryons containing two heavy quarks in bag model}},
  }{}\href{https://doi.org/10.1103/PhysRevD.70.094004}{Phys.\ Rev.\
  \textbf{D70} (2004) 094004},
  \href{http://arxiv.org/abs/hep-ph/0403301}{{\normalfont\ttfamily
  arXiv:hep-ph/0403301}}\relax
\mciteBstWouldAddEndPuncttrue
\mciteSetBstMidEndSepPunct{\mcitedefaultmidpunct}
{\mcitedefaultendpunct}{\mcitedefaultseppunct}\relax
\EndOfBibitem
\bibitem{qm}
D.~Ebert, R.~N. Faustov, V.~O. Galkin, and A.~P. Martynenko,
  \ifthenelse{\boolean{articletitles}}{\emph{{Mass spectra of doubly heavy
  baryons in the relativistic quark model}},
  }{}\href{https://doi.org/10.1103/PhysRevD.66.014008}{Phys.\ Rev.\
  \textbf{D66} (2002) 014008},
  \href{http://arxiv.org/abs/hep-ph/0201217}{{\normalfont\ttfamily
  arXiv:hep-ph/0201217}}\relax
\mciteBstWouldAddEndPuncttrue
\mciteSetBstMidEndSepPunct{\mcitedefaultmidpunct}
{\mcitedefaultendpunct}{\mcitedefaultseppunct}\relax
\EndOfBibitem
\bibitem{mass1}
S.~Fleck and J.-M. Richard, \ifthenelse{\boolean{articletitles}}{\emph{{Baryons
  with double charm}}, }{}\href{https://doi.org/10.1143/PTP.82.760}{Prog.\
  Theor.\ Phys.\  \textbf{82} (1989) 760}\relax
\mciteBstWouldAddEndPuncttrue
\mciteSetBstMidEndSepPunct{\mcitedefaultmidpunct}
{\mcitedefaultendpunct}{\mcitedefaultseppunct}\relax
\EndOfBibitem
\bibitem{mass2}
B.~O. Kerbikov, M.~I. Polikarpov, and L.~V. Shevchenko,
  \ifthenelse{\boolean{articletitles}}{\emph{{Multiquark masses and wave
  functions through a modified green function Monte Carlo method}},
  }{}\href{https://doi.org/10.1016/0550-3213(90)90016-7}{Nucl.\ Phys.\
  \textbf{B331} (1990) 19}\relax
\mciteBstWouldAddEndPuncttrue
\mciteSetBstMidEndSepPunct{\mcitedefaultmidpunct}
{\mcitedefaultendpunct}{\mcitedefaultseppunct}\relax
\EndOfBibitem
\bibitem{mass3}
S.~Chernyshev, M.~A. Nowak, and I.~Zahed,
  \ifthenelse{\boolean{articletitles}}{\emph{{Heavy hadrons and QCD
  instantons}}, }{}\href{https://doi.org/10.1103/PhysRevD.53.5176}{Phys.\ Rev.\
   \textbf{D53} (1996) 5176}\relax
\mciteBstWouldAddEndPuncttrue
\mciteSetBstMidEndSepPunct{\mcitedefaultmidpunct}
{\mcitedefaultendpunct}{\mcitedefaultseppunct}\relax
\EndOfBibitem
\bibitem{mass4}
S.~S. Gershtein, V.~V. Kiselev, A.~K. Likhoded, and A.~I. Onishchenko,
  \ifthenelse{\boolean{articletitles}}{\emph{{Spectroscopy of doubly heavy
  baryons}}, }{}\href{https://doi.org/10.1103/PhysRevD.62.054021}{Phys.\ Rev.\
  \textbf{D62} (2000) 054021}\relax
\mciteBstWouldAddEndPuncttrue
\mciteSetBstMidEndSepPunct{\mcitedefaultmidpunct}
{\mcitedefaultendpunct}{\mcitedefaultseppunct}\relax
\EndOfBibitem
\bibitem{mass5}
C.~Itoh, T.~Minamikawa, K.~Miura, and T.~Watanabe,
  \ifthenelse{\boolean{articletitles}}{\emph{{Doubly charmed baryon masses and
  quark wave functions in baryons}},
  }{}\href{https://doi.org/10.1103/PhysRevD.61.057502}{Phys.\ Rev.\
  \textbf{D61} (2000) 057502}\relax
\mciteBstWouldAddEndPuncttrue
\mciteSetBstMidEndSepPunct{\mcitedefaultmidpunct}
{\mcitedefaultendpunct}{\mcitedefaultseppunct}\relax
\EndOfBibitem
\bibitem{lifetime3}
V.~V. Kiselev and A.~K. Likhoded,
  \ifthenelse{\boolean{articletitles}}{\emph{{Baryons with two heavy quarks}},
  }{}\href{https://doi.org/10.1070/PU2002v045n05ABEH000958}{Phys.\ Usp.\
  \textbf{45} (2002) 455},
  \href{http://arxiv.org/abs/hep-ph/0103169}{{\normalfont\ttfamily
  arXiv:hep-ph/0103169}}\relax
\mciteBstWouldAddEndPuncttrue
\mciteSetBstMidEndSepPunct{\mcitedefaultmidpunct}
{\mcitedefaultendpunct}{\mcitedefaultseppunct}\relax
\EndOfBibitem
\bibitem{mass6}
N.~Mathur, R.~Lewis, and R.~M. Woloshyn,
  \ifthenelse{\boolean{articletitles}}{\emph{{Charmed and bottom baryons from
  lattice nonrelativistic QCD}},
  }{}\href{https://doi.org/10.1103/PhysRevD.66.014502}{Phys.\ Rev.\
  \textbf{D66} (2002) 014502}\relax
\mciteBstWouldAddEndPuncttrue
\mciteSetBstMidEndSepPunct{\mcitedefaultmidpunct}
{\mcitedefaultendpunct}{\mcitedefaultseppunct}\relax
\EndOfBibitem
\bibitem{mass7}
W.~Roberts and M.~Pervin, \ifthenelse{\boolean{articletitles}}{\emph{{Heavy
  baryons in a quark model}},
  }{}\href{https://doi.org/10.1142/S0217751X08041219}{Int.\ J.\ Mod.\ Phys.\
  \textbf{A23} (2008) 2817},
  \href{http://arxiv.org/abs/0711.2492}{{\normalfont\ttfamily
  arXiv:0711.2492}}\relax
\mciteBstWouldAddEndPuncttrue
\mciteSetBstMidEndSepPunct{\mcitedefaultmidpunct}
{\mcitedefaultendpunct}{\mcitedefaultseppunct}\relax
\EndOfBibitem
\bibitem{mass8}
A.~Valcarce, H.~Garcilazo, and J.~Vijande,
  \ifthenelse{\boolean{articletitles}}{\emph{{Towards an understanding of heavy
  baryon spectroscopy}},
  }{}\href{https://doi.org/10.1140/epja/i2008-10616-4}{Eur.\ Phys.\ J.\
  \textbf{A37} (2008) 217}\relax
\mciteBstWouldAddEndPuncttrue
\mciteSetBstMidEndSepPunct{\mcitedefaultmidpunct}
{\mcitedefaultendpunct}{\mcitedefaultseppunct}\relax
\EndOfBibitem
\bibitem{mass10}
T.~M. Aliev, K.~Azizi, and M.~Savci,
  \ifthenelse{\boolean{articletitles}}{\emph{{Doubly heavy spin--1/2 baryon
  spectrum in QCD}},
  }{}\href{https://doi.org/10.1016/j.nuclphysa.2012.09.009}{Nucl.\ Phys.\
  \textbf{A895} (2012) 59},
  \href{http://arxiv.org/abs/1205.2873}{{\normalfont\ttfamily
  arXiv:1205.2873}}\relax
\mciteBstWouldAddEndPuncttrue
\mciteSetBstMidEndSepPunct{\mcitedefaultmidpunct}
{\mcitedefaultendpunct}{\mcitedefaultseppunct}\relax
\EndOfBibitem
\bibitem{mass11}
PACS-CS collaboration, Y.~Namekawa {\em et~al.},
  \ifthenelse{\boolean{articletitles}}{\emph{{Charmed baryons at the physical
  point in 2+1 flavor lattice QCD}},
  }{}\href{https://doi.org/10.1103/PhysRevD.87.094512}{Phys.\ Rev.\
  \textbf{D87} (2013) 094512},
  \href{http://arxiv.org/abs/1301.4743}{{\normalfont\ttfamily
  arXiv:1301.4743}}\relax
\mciteBstWouldAddEndPuncttrue
\mciteSetBstMidEndSepPunct{\mcitedefaultmidpunct}
{\mcitedefaultendpunct}{\mcitedefaultseppunct}\relax
\EndOfBibitem
\bibitem{mass12}
Z.-F. Sun, Z.-W. Liu, X.~Liu, and S.-L. Zhu,
  \ifthenelse{\boolean{articletitles}}{\emph{{Masses and axial currents of the
  doubly charmed baryons}},
  }{}\href{https://doi.org/10.1103/PhysRevD.91.094030}{Phys.\ Rev.\
  \textbf{D91} (2015) 094030},
  \href{http://arxiv.org/abs/1411.2117}{{\normalfont\ttfamily
  arXiv:1411.2117}}\relax
\mciteBstWouldAddEndPuncttrue
\mciteSetBstMidEndSepPunct{\mcitedefaultmidpunct}
{\mcitedefaultendpunct}{\mcitedefaultseppunct}\relax
\EndOfBibitem
\bibitem{mass13}
M.~Padmanath, R.~G. Edwards, N.~Mathur, and M.~Peardon,
  \ifthenelse{\boolean{articletitles}}{\emph{{Spectroscopy of doubly-charmed
  baryons from lattice QCD}},
  }{}\href{https://doi.org/10.1103/PhysRevD.91.094502}{Phys.\ Rev.\
  \textbf{D91} (2015) 094502},
  \href{http://arxiv.org/abs/1502.01845}{{\normalfont\ttfamily
  arXiv:1502.01845}}\relax
\mciteBstWouldAddEndPuncttrue
\mciteSetBstMidEndSepPunct{\mcitedefaultmidpunct}
{\mcitedefaultendpunct}{\mcitedefaultseppunct}\relax
\EndOfBibitem
\bibitem{mass14}
P.~Pérez-Rubio, S.~Collins, and G.~S. Bali,
  \ifthenelse{\boolean{articletitles}}{\emph{{Charmed baryon spectroscopy and
  light flavor symmetry from lattice QCD}},
  }{}\href{https://doi.org/10.1103/PhysRevD.92.034504}{Phys.\ Rev.\
  \textbf{D92} (2015) 034504},
  \href{http://arxiv.org/abs/1503.08440}{{\normalfont\ttfamily
  arXiv:1503.08440}}\relax
\mciteBstWouldAddEndPuncttrue
\mciteSetBstMidEndSepPunct{\mcitedefaultmidpunct}
{\mcitedefaultendpunct}{\mcitedefaultseppunct}\relax
\EndOfBibitem
\bibitem{mass15}
K.-W. Wei, B.~Chen, and X.-H. Guo,
  \ifthenelse{\boolean{articletitles}}{\emph{{Masses of doubly and triply
  charmed baryons}},
  }{}\href{https://doi.org/10.1103/PhysRevD.92.076008}{Phys.\ Rev.\
  \textbf{D92} (2015) 076008},
  \href{http://arxiv.org/abs/1503.05184}{{\normalfont\ttfamily
  arXiv:1503.05184}}\relax
\mciteBstWouldAddEndPuncttrue
\mciteSetBstMidEndSepPunct{\mcitedefaultmidpunct}
{\mcitedefaultendpunct}{\mcitedefaultseppunct}\relax
\EndOfBibitem
\bibitem{mass16}
Z.-F. Sun and M.~J. Vicente~Vacas,
  \ifthenelse{\boolean{articletitles}}{\emph{{Masses of doubly charmed baryons
  in the extended on-mass-shell renormalization scheme}},
  }{}\href{https://doi.org/10.1103/PhysRevD.93.094002}{Phys.\ Rev.\
  \textbf{D93} (2016) 094002},
  \href{http://arxiv.org/abs/1602.04714}{{\normalfont\ttfamily
  arXiv:1602.04714}}\relax
\mciteBstWouldAddEndPuncttrue
\mciteSetBstMidEndSepPunct{\mcitedefaultmidpunct}
{\mcitedefaultendpunct}{\mcitedefaultseppunct}\relax
\EndOfBibitem
\bibitem{mass17}
C.~Alexandrou and C.~Kallidonis,
  \ifthenelse{\boolean{articletitles}}{\emph{{Low-lying baryon masses using
  $N_f=2$ twisted mass clover-improved fermions directly at the physical pion
  mass}}, }{}\href{https://doi.org/10.1103/PhysRevD.96.034511}{Phys.\ Rev.\
  \textbf{D96} (2017) 034511},
  \href{http://arxiv.org/abs/1704.02647}{{\normalfont\ttfamily
  arXiv:1704.02647}}\relax
\mciteBstWouldAddEndPuncttrue
\mciteSetBstMidEndSepPunct{\mcitedefaultmidpunct}
{\mcitedefaultendpunct}{\mcitedefaultseppunct}\relax
\EndOfBibitem
\bibitem{mass18}
Y.~Liu and I.~Zahed, \ifthenelse{\boolean{articletitles}}{\emph{{Heavy and
  strange holographic baryons}},
  }{}\href{https://doi.org/10.1103/PhysRevD.96.056027}{Phys.\ Rev.\
  \textbf{D96} (2017) 056027},
  \href{http://arxiv.org/abs/1705.01397}{{\normalfont\ttfamily
  arXiv:1705.01397}}\relax
\mciteBstWouldAddEndPuncttrue
\mciteSetBstMidEndSepPunct{\mcitedefaultmidpunct}
{\mcitedefaultendpunct}{\mcitedefaultseppunct}\relax
\EndOfBibitem
\bibitem{isospinsplit3}
C.-W. Hwang and C.-H. Chung,
  \ifthenelse{\boolean{articletitles}}{\emph{{Isospin mass splittings of heavy
  baryons in heavy quark symmetry}},
  }{}\href{https://doi.org/10.1103/PhysRevD.78.073013}{Phys.\ Rev.\
  \textbf{D78} (2008) 073013}\relax
\mciteBstWouldAddEndPuncttrue
\mciteSetBstMidEndSepPunct{\mcitedefaultmidpunct}
{\mcitedefaultendpunct}{\mcitedefaultseppunct}\relax
\EndOfBibitem
\bibitem{isospinsplit}
S.~J. Brodsky, F.~Guo, C.~Hanhart, and U.~Mei{\ss}ner,
  \ifthenelse{\boolean{articletitles}}{\emph{{Isospin splittings of doubly
  heavy baryons}},
  }{}\href{https://doi.org/10.1016/j.physletb.2011.03.014}{Phys.\ Lett.\
  \textbf{B698} (2011) 251},
  \href{http://arxiv.org/abs/1101.1983}{{\normalfont\ttfamily
  arXiv:1101.1983}}\relax
\mciteBstWouldAddEndPuncttrue
\mciteSetBstMidEndSepPunct{\mcitedefaultmidpunct}
{\mcitedefaultendpunct}{\mcitedefaultseppunct}\relax
\EndOfBibitem
\bibitem{isospinsplit2}
M.~Karliner and J.~L. Rosner,
  \ifthenelse{\boolean{articletitles}}{\emph{{Isospin splittings in baryons
  with two heavy quarks}},
  }{}\href{https://doi.org/10.1103/PhysRevD.96.033004}{Phys.\ Rev.\
  \textbf{D96} (2017) 033004}\relax
\mciteBstWouldAddEndPuncttrue
\mciteSetBstMidEndSepPunct{\mcitedefaultmidpunct}
{\mcitedefaultendpunct}{\mcitedefaultseppunct}\relax
\EndOfBibitem
\bibitem{lifetime0}
B.~Guberina, B.~Melić, and H.~Štefančić,
  \ifthenelse{\boolean{articletitles}}{\emph{{Inclusive decays and lifetimes of
  doubly charmed baryons}},
  }{}\href{https://doi.org/10.1007/s100529900039}{Eur.\ Phys.\ J.\  \textbf{C9}
  (1999) 213}, \href{http://arxiv.org/abs/hep-ph/9901323}{{\normalfont\ttfamily
  arXiv:hep-ph/9901323}}\relax
\mciteBstWouldAddEndPuncttrue
\mciteSetBstMidEndSepPunct{\mcitedefaultmidpunct}
{\mcitedefaultendpunct}{\mcitedefaultseppunct}\relax
\EndOfBibitem
\bibitem{lifetime1}
V.~V. Kiselev, A.~K. Likhoded, and A.~I. Onishchenko,
  \ifthenelse{\boolean{articletitles}}{\emph{{Lifetimes of doubly charmed
  baryons: \Xiccp and \Xiccpp}},
  }{}\href{https://doi.org/10.1103/PhysRevD.60.014007}{Phys.\ Rev.\
  \textbf{D60} (1999) 014007},
  \href{http://arxiv.org/abs/hep-ph/9807354}{{\normalfont\ttfamily
  arXiv:hep-ph/9807354}}\relax
\mciteBstWouldAddEndPuncttrue
\mciteSetBstMidEndSepPunct{\mcitedefaultmidpunct}
{\mcitedefaultendpunct}{\mcitedefaultseppunct}\relax
\EndOfBibitem
\bibitem{lifetime2}
A.~K. Likhoded and A.~I. Onishchenko,
  \ifthenelse{\boolean{articletitles}}{\emph{{Lifetimes of doubly heavy
  baryons}},
  }{}\href{http://arxiv.org/abs/hep-ph/9912425}{{\normalfont\ttfamily
  arXiv:hep-ph/9912425}}\relax
\mciteBstWouldAddEndPuncttrue
\mciteSetBstMidEndSepPunct{\mcitedefaultmidpunct}
{\mcitedefaultendpunct}{\mcitedefaultseppunct}\relax
\EndOfBibitem
\bibitem{lifetime4}
C.~Chang, T.~Li, X.~Li, and Y.~Wang,
  \ifthenelse{\boolean{articletitles}}{\emph{{Lifetime of doubly charmed
  baryons}}, }{}\href{https://doi.org/10.1088/0253-6102/49/4/38}{Commun.\
  Theor.\ Phys.\  \textbf{49} (2008) 993},
  \href{http://arxiv.org/abs/0704.0016}{{\normalfont\ttfamily
  arXiv:0704.0016}}\relax
\mciteBstWouldAddEndPuncttrue
\mciteSetBstMidEndSepPunct{\mcitedefaultmidpunct}
{\mcitedefaultendpunct}{\mcitedefaultseppunct}\relax
\EndOfBibitem
\bibitem{lifetime5}
A.~V. Berezhnoy and A.~K. Likhoded,
  \ifthenelse{\boolean{articletitles}}{\emph{{Doubly heavy baryons}},
  }{}\href{https://doi.org/10.1134/S1063778816010087}{Phys.\ Atom.\ Nucl.\
  \textbf{79} (2016) 260}\relax
\mciteBstWouldAddEndPuncttrue
\mciteSetBstMidEndSepPunct{\mcitedefaultmidpunct}
{\mcitedefaultendpunct}{\mcitedefaultseppunct}\relax
\EndOfBibitem
\bibitem{lifetime6}
H.-Y. Cheng and Y.-L. Shi,
  \ifthenelse{\boolean{articletitles}}{\emph{{Lifetimes of Doubly Charmed
  Baryons}}, }{}\href{https://doi.org/10.1103/PhysRevD.98.113005}{Phys.\ Rev.\
  \textbf{D98} (2018) 113005},
  \href{http://arxiv.org/abs/1809.08102}{{\normalfont\ttfamily
  arXiv:1809.08102}}\relax
\mciteBstWouldAddEndPuncttrue
\mciteSetBstMidEndSepPunct{\mcitedefaultmidpunct}
{\mcitedefaultendpunct}{\mcitedefaultseppunct}\relax
\EndOfBibitem
\bibitem{bftheory2}
Y.-J. Shi, W.~Wang, Y.~Xing, and J.~Xu,
  \ifthenelse{\boolean{articletitles}}{\emph{{Weak Decays of Doubly Heavy
  Baryons: Multi-body Decay Channels}},
  }{}\href{https://doi.org/10.1140/epjc/s10052-018-5532-7}{Eur.\ Phys.\ J.\
  \textbf{C78} (2018) 56},
  \href{http://arxiv.org/abs/1712.03830}{{\normalfont\ttfamily
  arXiv:1712.03830}}\relax
\mciteBstWouldAddEndPuncttrue
\mciteSetBstMidEndSepPunct{\mcitedefaultmidpunct}
{\mcitedefaultendpunct}{\mcitedefaultseppunct}\relax
\EndOfBibitem
\bibitem{bftheory3}
L.-J. Jiang, B.~He, and R.-H. Li,
  \ifthenelse{\boolean{articletitles}}{\emph{{Weak decays of doubly heavy
  baryons: $\mathcal{B}_{cc}\rightarrow \mathcal{B}_c V$}},
  }{}\href{https://doi.org/10.1140/epjc/s10052-018-6445-1}{Eur.\ Phys.\ J.\
  \textbf{C78} (2018) 961},
  \href{http://arxiv.org/abs/1810.00541}{{\normalfont\ttfamily
  arXiv:1810.00541}}\relax
\mciteBstWouldAddEndPuncttrue
\mciteSetBstMidEndSepPunct{\mcitedefaultmidpunct}
{\mcitedefaultendpunct}{\mcitedefaultseppunct}\relax
\EndOfBibitem
\bibitem{hadronprod}
C.-H. Chang, C.-F. Qiao, J.-X. Wang, and X.-G. Wu,
  \ifthenelse{\boolean{articletitles}}{\emph{{Estimate of the hadronic
  production of the doubly charmed baryon ${\ensuremath{\Xi}}_{cc}$ in the
  general-mass variable-flavor-number scheme}},
  }{}\href{https://doi.org/10.1103/PhysRevD.73.094022}{Phys.\ Rev.\
  \textbf{D73} (2006) 094022}\relax
\mciteBstWouldAddEndPuncttrue
\mciteSetBstMidEndSepPunct{\mcitedefaultmidpunct}
{\mcitedefaultendpunct}{\mcitedefaultseppunct}\relax
\EndOfBibitem
\bibitem{LHCb-DP-2008-001}
LHCb collaboration, A.~A. Alves~Jr.\ {\em et~al.},
  \ifthenelse{\boolean{articletitles}}{\emph{{The \lhcb detector at the LHC}},
  }{}\href{https://doi.org/10.1088/1748-0221/3/08/S08005}{JINST \textbf{3}
  (2008) S08005}\relax
\mciteBstWouldAddEndPuncttrue
\mciteSetBstMidEndSepPunct{\mcitedefaultmidpunct}
{\mcitedefaultendpunct}{\mcitedefaultseppunct}\relax
\EndOfBibitem
\bibitem{LHCb-DP-2014-002}
LHCb collaboration, R.~Aaij {\em et~al.},
  \ifthenelse{\boolean{articletitles}}{\emph{{LHCb detector performance}},
  }{}\href{https://doi.org/10.1142/S0217751X15300227}{Int.\ J.\ Mod.\ Phys.\
  \textbf{A30} (2015) 1530022},
  \href{http://arxiv.org/abs/1412.6352}{{\normalfont\ttfamily
  arXiv:1412.6352}}\relax
\mciteBstWouldAddEndPuncttrue
\mciteSetBstMidEndSepPunct{\mcitedefaultmidpunct}
{\mcitedefaultendpunct}{\mcitedefaultseppunct}\relax
\EndOfBibitem
\bibitem{LHCb-DP-2014-001}
R.~Aaij {\em et~al.}, \ifthenelse{\boolean{articletitles}}{\emph{{Performance
  of the LHCb Vertex Locator}},
  }{}\href{https://doi.org/10.1088/1748-0221/9/09/P09007}{JINST \textbf{9}
  (2014) P09007}, \href{http://arxiv.org/abs/1405.7808}{{\normalfont\ttfamily
  arXiv:1405.7808}}\relax
\mciteBstWouldAddEndPuncttrue
\mciteSetBstMidEndSepPunct{\mcitedefaultmidpunct}
{\mcitedefaultendpunct}{\mcitedefaultseppunct}\relax
\EndOfBibitem
\bibitem{LHCb-DP-2017-001}
P.~d'Argent {\em et~al.}, \ifthenelse{\boolean{articletitles}}{\emph{{Improved
  performance of the LHCb Outer Tracker in LHC Run 2}},
  }{}\href{https://doi.org/10.1088/1748-0221/12/11/P11016}{JINST \textbf{12}
  (2017) P11016}, \href{http://arxiv.org/abs/1708.00819}{{\normalfont\ttfamily
  arXiv:1708.00819}}\relax
\mciteBstWouldAddEndPuncttrue
\mciteSetBstMidEndSepPunct{\mcitedefaultmidpunct}
{\mcitedefaultendpunct}{\mcitedefaultseppunct}\relax
\EndOfBibitem
\bibitem{LHCb-DP-2012-003}
M.~Adinolfi {\em et~al.},
  \ifthenelse{\boolean{articletitles}}{\emph{{Performance of the \lhcb RICH
  detector at the LHC}},
  }{}\href{https://doi.org/10.1140/epjc/s10052-013-2431-9}{Eur.\ Phys.\ J.\
  \textbf{C73} (2013) 2431},
  \href{http://arxiv.org/abs/1211.6759}{{\normalfont\ttfamily
  arXiv:1211.6759}}\relax
\mciteBstWouldAddEndPuncttrue
\mciteSetBstMidEndSepPunct{\mcitedefaultmidpunct}
{\mcitedefaultendpunct}{\mcitedefaultseppunct}\relax
\EndOfBibitem
\bibitem{LHCb-DP-2020-001}
C.~Abellan~Beteta {\em et~al.},
  \ifthenelse{\boolean{articletitles}}{\emph{{Calibration and performance of
  the LHCb calorimeters in Run 1 and 2 at the LHC}},
  }{}\href{http://arxiv.org/abs/2008.11556}{{\normalfont\ttfamily
  arXiv:2008.11556}}, {submitted to JINST}\relax
\mciteBstWouldAddEndPuncttrue
\mciteSetBstMidEndSepPunct{\mcitedefaultmidpunct}
{\mcitedefaultendpunct}{\mcitedefaultseppunct}\relax
\EndOfBibitem
\bibitem{LHCb-DP-2012-002}
A.~A. Alves~Jr.\ {\em et~al.},
  \ifthenelse{\boolean{articletitles}}{\emph{{Performance of the LHCb muon
  system}}, }{}\href{https://doi.org/10.1088/1748-0221/8/02/P02022}{JINST
  \textbf{8} (2013) P02022},
  \href{http://arxiv.org/abs/1211.1346}{{\normalfont\ttfamily
  arXiv:1211.1346}}\relax
\mciteBstWouldAddEndPuncttrue
\mciteSetBstMidEndSepPunct{\mcitedefaultmidpunct}
{\mcitedefaultendpunct}{\mcitedefaultseppunct}\relax
\EndOfBibitem
\bibitem{LHCb-DP-2012-004}
R.~Aaij {\em et~al.}, \ifthenelse{\boolean{articletitles}}{\emph{{The \lhcb
  trigger and its performance in 2011}},
  }{}\href{https://doi.org/10.1088/1748-0221/8/04/P04022}{JINST \textbf{8}
  (2013) P04022}, \href{http://arxiv.org/abs/1211.3055}{{\normalfont\ttfamily
  arXiv:1211.3055}}\relax
\mciteBstWouldAddEndPuncttrue
\mciteSetBstMidEndSepPunct{\mcitedefaultmidpunct}
{\mcitedefaultendpunct}{\mcitedefaultseppunct}\relax
\EndOfBibitem
\bibitem{LHCb-PROC-2015-011}
G.~Dujany and B.~Storaci, \ifthenelse{\boolean{articletitles}}{\emph{{Real-time
  alignment and calibration of the LHCb Detector in Run II}},
  }{}\href{https://doi.org/10.1088/1742-6596/664/8/082010}{J.\ Phys.\ Conf.\
  Ser.\  \textbf{664} (2015) 082010}\relax
\mciteBstWouldAddEndPuncttrue
\mciteSetBstMidEndSepPunct{\mcitedefaultmidpunct}
{\mcitedefaultendpunct}{\mcitedefaultseppunct}\relax
\EndOfBibitem
\bibitem{LHCb-DP-2016-001}
R.~Aaij {\em et~al.}, \ifthenelse{\boolean{articletitles}}{\emph{{Tesla: an
  application for real-time data analysis in High Energy Physics}},
  }{}\href{https://doi.org/10.1016/j.cpc.2016.07.022}{Comput.\ Phys.\ Commun.\
  \textbf{208} (2016) 35},
  \href{http://arxiv.org/abs/1604.05596}{{\normalfont\ttfamily
  arXiv:1604.05596}}\relax
\mciteBstWouldAddEndPuncttrue
\mciteSetBstMidEndSepPunct{\mcitedefaultmidpunct}
{\mcitedefaultendpunct}{\mcitedefaultseppunct}\relax
\EndOfBibitem
\bibitem{LHCb-PAPER-2012-048}
LHCb collaboration, R.~Aaij {\em et~al.},
  \ifthenelse{\boolean{articletitles}}{\emph{{Measurements of the \Lb, \Xibm,
  and \Omegab baryon masses}},
  }{}\href{https://doi.org/10.1103/PhysRevLett.110.182001}{Phys.\ Rev.\ Lett.\
  \textbf{110} (2013) 182001},
  \href{http://arxiv.org/abs/1302.1072}{{\normalfont\ttfamily
  arXiv:1302.1072}}\relax
\mciteBstWouldAddEndPuncttrue
\mciteSetBstMidEndSepPunct{\mcitedefaultmidpunct}
{\mcitedefaultendpunct}{\mcitedefaultseppunct}\relax
\EndOfBibitem
\bibitem{LHCb-PAPER-2013-011}
LHCb collaboration, R.~Aaij {\em et~al.},
  \ifthenelse{\boolean{articletitles}}{\emph{{Precision measurement of \D meson
  mass differences}}, }{}\href{https://doi.org/10.1007/JHEP06(2013)065}{JHEP
  \textbf{06} (2013) 065},
  \href{http://arxiv.org/abs/1304.6865}{{\normalfont\ttfamily
  arXiv:1304.6865}}\relax
\mciteBstWouldAddEndPuncttrue
\mciteSetBstMidEndSepPunct{\mcitedefaultmidpunct}
{\mcitedefaultendpunct}{\mcitedefaultseppunct}\relax
\EndOfBibitem
\bibitem{Sjostrand:2007gs}
T.~Sj\"{o}strand, S.~Mrenna, and P.~Skands,
  \ifthenelse{\boolean{articletitles}}{\emph{{A brief introduction to PYTHIA
  8.1}}, }{}\href{https://doi.org/10.1016/j.cpc.2008.01.036}{Comput.\ Phys.\
  Commun.\  \textbf{178} (2008) 852},
  \href{http://arxiv.org/abs/0710.3820}{{\normalfont\ttfamily
  arXiv:0710.3820}}\relax
\mciteBstWouldAddEndPuncttrue
\mciteSetBstMidEndSepPunct{\mcitedefaultmidpunct}
{\mcitedefaultendpunct}{\mcitedefaultseppunct}\relax
\EndOfBibitem
\bibitem{Sjostrand:2006za}
T.~Sj\"{o}strand, S.~Mrenna, and P.~Skands,
  \ifthenelse{\boolean{articletitles}}{\emph{{PYTHIA 6.4 physics and manual}},
  }{}\href{https://doi.org/10.1088/1126-6708/2006/05/026}{JHEP \textbf{05}
  (2006) 026}, \href{http://arxiv.org/abs/hep-ph/0603175}{{\normalfont\ttfamily
  arXiv:hep-ph/0603175}}\relax
\mciteBstWouldAddEndPuncttrue
\mciteSetBstMidEndSepPunct{\mcitedefaultmidpunct}
{\mcitedefaultendpunct}{\mcitedefaultseppunct}\relax
\EndOfBibitem
\bibitem{LHCb-PROC-2010-056}
I.~Belyaev {\em et~al.}, \ifthenelse{\boolean{articletitles}}{\emph{{Handling
  of the generation of primary events in Gauss, the LHCb simulation
  framework}}, }{}\href{https://doi.org/10.1088/1742-6596/331/3/032047}{J.\
  Phys.\ Conf.\ Ser.\  \textbf{331} (2011) 032047}\relax
\mciteBstWouldAddEndPuncttrue
\mciteSetBstMidEndSepPunct{\mcitedefaultmidpunct}
{\mcitedefaultendpunct}{\mcitedefaultseppunct}\relax
\EndOfBibitem
\bibitem{genxicc}
C.-H. Chang, J.-X. Wang, and X.-G. Wu,
  \ifthenelse{\boolean{articletitles}}{\emph{{GENXICC2.0: An upgraded version
  of the generator for hadronic production of double heavy baryons \Xiccunsign,
  \Xibc and \Xibb}},
  }{}\href{https://doi.org/10.1016/j.cpc.2010.02.008}{Comput.\ Phys.\ Commun.\
  \textbf{181} (2010) 1144},
  \href{http://arxiv.org/abs/0910.4462}{{\normalfont\ttfamily
  arXiv:0910.4462}}\relax
\mciteBstWouldAddEndPuncttrue
\mciteSetBstMidEndSepPunct{\mcitedefaultmidpunct}
{\mcitedefaultendpunct}{\mcitedefaultseppunct}\relax
\EndOfBibitem
\bibitem{Lange:2001uf}
D.~J. Lange, \ifthenelse{\boolean{articletitles}}{\emph{{The EvtGen particle
  decay simulation package}},
  }{}\href{https://doi.org/10.1016/S0168-9002(01)00089-4}{Nucl.\ Instrum.\
  Meth.\  \textbf{A462} (2001) 152}\relax
\mciteBstWouldAddEndPuncttrue
\mciteSetBstMidEndSepPunct{\mcitedefaultmidpunct}
{\mcitedefaultendpunct}{\mcitedefaultseppunct}\relax
\EndOfBibitem
\bibitem{davidson2015photos}
N.~Davidson, T.~Przedzinski, and Z.~Was,
  \ifthenelse{\boolean{articletitles}}{\emph{{PHOTOS interface in C++:
  Technical and physics documentation}},
  }{}\href{https://doi.org/https://doi.org/10.1016/j.cpc.2015.09.013}{Comp.\
  Phys.\ Comm.\  \textbf{199} (2016) 86},
  \href{http://arxiv.org/abs/1011.0937}{{\normalfont\ttfamily
  arXiv:1011.0937}}\relax
\mciteBstWouldAddEndPuncttrue
\mciteSetBstMidEndSepPunct{\mcitedefaultmidpunct}
{\mcitedefaultendpunct}{\mcitedefaultseppunct}\relax
\EndOfBibitem
\bibitem{Allison:2006ve}
Geant4 collaboration, J.~Allison {\em et~al.},
  \ifthenelse{\boolean{articletitles}}{\emph{{Geant4 developments and
  applications}}, }{}\href{https://doi.org/10.1109/TNS.2006.869826}{IEEE
  Trans.\ Nucl.\ Sci.\  \textbf{53} (2006) 270}\relax
\mciteBstWouldAddEndPuncttrue
\mciteSetBstMidEndSepPunct{\mcitedefaultmidpunct}
{\mcitedefaultendpunct}{\mcitedefaultseppunct}\relax
\EndOfBibitem
\bibitem{Agostinelli:2002hh}
Geant4 collaboration, S.~Agostinelli {\em et~al.},
  \ifthenelse{\boolean{articletitles}}{\emph{{Geant4: A simulation toolkit}},
  }{}\href{https://doi.org/10.1016/S0168-9002(03)01368-8}{Nucl.\ Instrum.\
  Meth.\  \textbf{A506} (2003) 250}\relax
\mciteBstWouldAddEndPuncttrue
\mciteSetBstMidEndSepPunct{\mcitedefaultmidpunct}
{\mcitedefaultendpunct}{\mcitedefaultseppunct}\relax
\EndOfBibitem
\bibitem{LHCb-PROC-2011-006}
M.~Clemencic {\em et~al.}, \ifthenelse{\boolean{articletitles}}{\emph{{The
  \lhcb simulation application, Gauss: Design, evolution and experience}},
  }{}\href{https://doi.org/10.1088/1742-6596/331/3/032023}{J.\ Phys.\ Conf.\
  Ser.\  \textbf{331} (2011) 032023}\relax
\mciteBstWouldAddEndPuncttrue
\mciteSetBstMidEndSepPunct{\mcitedefaultmidpunct}
{\mcitedefaultendpunct}{\mcitedefaultseppunct}\relax
\EndOfBibitem
\bibitem{PDG2019}
Particle Data Group, M.~Tanabashi {\em et~al.},
  \ifthenelse{\boolean{articletitles}}{\emph{{\href{http://pdg.lbl.gov/}{Review
  of particle physics}}},
  }{}\href{https://doi.org/10.1103/PhysRevD.98.030001}{Phys.\ Rev.\
  \textbf{D98} (2018) 030001}, and {\href{http://pdglive.lbl.gov/}{2019
  update}}\relax
\mciteBstWouldAddEndPuncttrue
\mciteSetBstMidEndSepPunct{\mcitedefaultmidpunct}
{\mcitedefaultendpunct}{\mcitedefaultseppunct}\relax
\EndOfBibitem
\bibitem{LHCb-DP-2019-001}
R.~Aaij {\em et~al.}, \ifthenelse{\boolean{articletitles}}{\emph{{Performance
  of the LHCb trigger and full real-time reconstruction in Run 2 of the LHC}},
  }{}\href{https://doi.org/10.1088/1748-0221/14/04/P04013}{JINST \textbf{14}
  (2019) P04013}, \href{http://arxiv.org/abs/1812.10790}{{\normalfont\ttfamily
  arXiv:1812.10790}}\relax
\mciteBstWouldAddEndPuncttrue
\mciteSetBstMidEndSepPunct{\mcitedefaultmidpunct}
{\mcitedefaultendpunct}{\mcitedefaultseppunct}\relax
\EndOfBibitem
\bibitem{BBDT}
V.~V. Gligorov and M.~Williams,
  \ifthenelse{\boolean{articletitles}}{\emph{{Efficient, reliable and fast
  high-level triggering using a bonsai boosted decision tree}},
  }{}\href{https://doi.org/10.1088/1748-0221/8/02/P02013}{JINST \textbf{8}
  (2013) P02013}, \href{http://arxiv.org/abs/1210.6861}{{\normalfont\ttfamily
  arXiv:1210.6861}}\relax
\mciteBstWouldAddEndPuncttrue
\mciteSetBstMidEndSepPunct{\mcitedefaultmidpunct}
{\mcitedefaultendpunct}{\mcitedefaultseppunct}\relax
\EndOfBibitem
\bibitem{LHCb-PROC-2015-018}
T.~Likhomanenko {\em et~al.}, \ifthenelse{\boolean{articletitles}}{\emph{{LHCb
  topological trigger reoptimization}},
  }{}\href{https://doi.org/10.1088/1742-6596/664/8/082025}{J.\ Phys.\ Conf.\
  Ser.\  \textbf{664} (2015) 082025}\relax
\mciteBstWouldAddEndPuncttrue
\mciteSetBstMidEndSepPunct{\mcitedefaultmidpunct}
{\mcitedefaultendpunct}{\mcitedefaultseppunct}\relax
\EndOfBibitem
\bibitem{Rogozhnikov:2016bdp}
A.~Rogozhnikov, \ifthenelse{\boolean{articletitles}}{\emph{{Reweighting with
  Boosted Decision Trees}},
  }{}\href{https://doi.org/10.1088/1742-6596/762/1/012036}{J.\ Phys.\ Conf.\
  Ser.\  \textbf{762} (2016) 012036},
  \href{http://arxiv.org/abs/1608.05806}{{\normalfont\ttfamily
  arXiv:1608.05806}}, \url{https://github.com/arogozhnikov/hep_ml}\relax
\mciteBstWouldAddEndPuncttrue
\mciteSetBstMidEndSepPunct{\mcitedefaultmidpunct}
{\mcitedefaultendpunct}{\mcitedefaultseppunct}\relax
\EndOfBibitem
\bibitem{Pivk:2004ty}
M.~Pivk and F.~R. Le~Diberder,
  \ifthenelse{\boolean{articletitles}}{\emph{{sPlot: A statistical tool to
  unfold data distributions}},
  }{}\href{https://doi.org/10.1016/j.nima.2005.08.106}{Nucl.\ Instrum.\ Meth.\
  \textbf{A555} (2005) 356},
  \href{http://arxiv.org/abs/physics/0402083}{{\normalfont\ttfamily
  arXiv:physics/0402083}}\relax
\mciteBstWouldAddEndPuncttrue
\mciteSetBstMidEndSepPunct{\mcitedefaultmidpunct}
{\mcitedefaultendpunct}{\mcitedefaultseppunct}\relax
\EndOfBibitem
\bibitem{Hocker:2007ht}
H.~Voss, A.~Hoecker, J.~Stelzer, and F.~Tegenfeldt,
  \ifthenelse{\boolean{articletitles}}{\emph{{TMVA - Toolkit for multivariate
  data analysis with ROOT}}, }{}\href{https://doi.org/10.22323/1.050.0040}{PoS
  \textbf{ACAT} (2007) 040}\relax
\mciteBstWouldAddEndPuncttrue
\mciteSetBstMidEndSepPunct{\mcitedefaultmidpunct}
{\mcitedefaultendpunct}{\mcitedefaultseppunct}\relax
\EndOfBibitem
\bibitem{Hulsbergen:2005pu}
W.~D. Hulsbergen, \ifthenelse{\boolean{articletitles}}{\emph{{Decay chain
  fitting with a Kalman filter}},
  }{}\href{https://doi.org/10.1016/j.nima.2005.06.078}{Nucl.\ Instrum.\ Meth.\
  \textbf{A552} (2005) 566},
  \href{http://arxiv.org/abs/physics/0503191}{{\normalfont\ttfamily
  arXiv:physics/0503191}}\relax
\mciteBstWouldAddEndPuncttrue
\mciteSetBstMidEndSepPunct{\mcitedefaultmidpunct}
{\mcitedefaultendpunct}{\mcitedefaultseppunct}\relax
\EndOfBibitem
\bibitem{mlp}
S.~Haykin, {\em {Neural networks: a comprehensive foundation, 2nd ed}}, New
  York: Macmillan College Publishing, 1998\relax
\mciteBstWouldAddEndPuncttrue
\mciteSetBstMidEndSepPunct{\mcitedefaultmidpunct}
{\mcitedefaultendpunct}{\mcitedefaultseppunct}\relax
\EndOfBibitem
\bibitem{Punzi:2003bu}
G.~Punzi, \ifthenelse{\boolean{articletitles}}{\emph{{Sensitivity of searches
  for new signals and its optimization}}, }{}eConf \textbf{C030908} (2003)
  MODT002, \href{http://arxiv.org/abs/physics/0308063}{{\normalfont\ttfamily
  arXiv:physics/0308063}}\relax
\mciteBstWouldAddEndPuncttrue
\mciteSetBstMidEndSepPunct{\mcitedefaultmidpunct}
{\mcitedefaultendpunct}{\mcitedefaultseppunct}\relax
\EndOfBibitem
\bibitem{asymptotic}
G.~Cowan, K.~Cranmer, E.~Gross, and O.~Vitells,
  \ifthenelse{\boolean{articletitles}}{\emph{{Asymptotic formulae for
  likelihood-based tests of new physics}},
  }{}\href{https://doi.org/10.1140/epjc/s10052-011-1554-0}{Eur.\ Phys.\ J.\
  \textbf{C71} (2011) 1554},
  \href{http://arxiv.org/abs/1007.1727}{{\normalfont\ttfamily
  arXiv:1007.1727}}, [Erratum: Eur. Phys. J. \textbf{C73} (2013) 2501]\relax
\mciteBstWouldAddEndPuncttrue
\mciteSetBstMidEndSepPunct{\mcitedefaultmidpunct}
{\mcitedefaultendpunct}{\mcitedefaultseppunct}\relax
\EndOfBibitem
\bibitem{Skwarnicki:1986xj}
T.~Skwarnicki, {\em {A study of the radiative cascade transitions between the
  Upsilon-prime and Upsilon resonances}}, PhD thesis, Institute of Nuclear
  Physics, Krakow, 1986,
  {\href{http://inspirehep.net/record/230779/}{DESY-F31-86-02}}\relax
\mciteBstWouldAddEndPuncttrue
\mciteSetBstMidEndSepPunct{\mcitedefaultmidpunct}
{\mcitedefaultendpunct}{\mcitedefaultseppunct}\relax
\EndOfBibitem
\bibitem{LHCb-DP-2013-002}
LHCb collaboration, R.~Aaij {\em et~al.},
  \ifthenelse{\boolean{articletitles}}{\emph{{Measurement of the track
  reconstruction efficiency at LHCb}},
  }{}\href{https://doi.org/10.1088/1748-0221/10/02/P02007}{JINST \textbf{10}
  (2015) P02007}, \href{http://arxiv.org/abs/1408.1251}{{\normalfont\ttfamily
  arXiv:1408.1251}}\relax
\mciteBstWouldAddEndPuncttrue
\mciteSetBstMidEndSepPunct{\mcitedefaultmidpunct}
{\mcitedefaultendpunct}{\mcitedefaultseppunct}\relax
\EndOfBibitem
\bibitem{clsmethod}
A.~L. Read, \ifthenelse{\boolean{articletitles}}{\emph{{Modified frequentist
  analysis of search results (The CL(s) method)}}, }{} in {\em {Workshop on
  Confidence Limits}}, 81--101, 2000\relax
\mciteBstWouldAddEndPuncttrue
\mciteSetBstMidEndSepPunct{\mcitedefaultmidpunct}
{\mcitedefaultendpunct}{\mcitedefaultseppunct}\relax
\EndOfBibitem
\bibitem{LHCb-TDR-012}
LHCb collaboration, \ifthenelse{\boolean{articletitles}}{\emph{{Framework TDR
  for the LHCb Upgrade: Technical Design Report}}, }{}
  \href{http://cdsweb.cern.ch/search?p=CERN-LHCC-2012-007&f=reportnumber&action_search=Search&c=LHCb}
  {CERN-LHCC-2012-007}, 2012\relax
\mciteBstWouldAddEndPuncttrue
\mciteSetBstMidEndSepPunct{\mcitedefaultmidpunct}
{\mcitedefaultendpunct}{\mcitedefaultseppunct}\relax
\EndOfBibitem
\end{mcitethebibliography}

\newpage
% LHCb collaboration author list
% Data extracted on September 8th, 2021 at 12:18pm for paper reference LHCb-PAPER-2021-019
\centerline
{\large\bf LHCb collaboration}
\begin
{flushleft}
\small
R.~Aaij$^{32}$,
A.S.W.~Abdelmotteleb$^{56}$,
C.~Abell{\'a}n~Beteta$^{50}$,
T.~Ackernley$^{60}$,
B.~Adeva$^{46}$,
M.~Adinolfi$^{54}$,
H.~Afsharnia$^{9}$,
C.~Agapopoulou$^{13}$,
C.A.~Aidala$^{86}$,
S.~Aiola$^{25}$
% ...

{\footnotesize \it

% ...
$^{9}$Universit{\'e} Clermont Auvergne, CNRS/IN2P3, LPC, Clermont-Ferrand, France\\
% ...
$^{13}$LPNHE, Sorbonne Universit{\'e}, Paris Diderot Sorbonne Paris Cit{\'e}, CNRS/IN2P3, Paris, France\\
% ...
$^{25}$INFN Sezione di Milano, Milano, Italy\\
% ...
$^{32}$Nikhef National Institute for Subatomic Physics, Amsterdam, Netherlands\\
% ...
$^{46}$Instituto Galego de F{\'\i}sica de Altas Enerx{\'\i}as (IGFAE), Universidade de Santiago de Compostela, Santiago de Compostela, Spain\\
% ...
$^{50}$Physik-Institut, Universit{\"a}t Z{\"u}rich, Z{\"u}rich, Switzerland\\
% ...
$^{54}$H.H. Wills Physics Laboratory, University of Bristol, Bristol, United Kingdom\\
% ...
$^{56}$Department of Physics, University of Warwick, Coventry, United Kingdom\\
% ...
$^{60}$Oliver Lodge Laboratory, University of Liverpool, Liverpool, United Kingdom\\
% ...
$^{68}$Syracuse University, Syracuse, NY, United States\\
% ...
$^{86}$University of Michigan, Ann Arbor, United States, associated to $^{68}$\\
\bigskip
% ...
\medskip
}
\end{flushleft}

\end{document}